\documentclass[12pt]{iopart}
\usepackage{iopams}

\input epsf.sty

\newlength{\TZ}
\newcommand{\BEQ}{\begin{equation}}     
\newcommand{\BEA}{\begin{eqnarray}}
\newcommand{\EEQ}{\end{equation}}       
\newcommand{\EEA}{\end{eqnarray}}
\newcommand{\D}{{\rm d}}                
\newcommand{\II}{{\rm i}}               
\newcommand{\wit}[1]{\widetilde{#1}}    
\newcommand{\wht}[1]{\widehat{#1}}      

\renewcommand{\vec}[1]{\boldsymbol{#1}} 

\newcommand{\appsection}[2]{\setcounter{equation}{0}\setcounter{subsection}{0}
\section*{Appendix #1. #2}
\renewcommand{\theequation}{#1\arabic{equation}}
              \renewcommand{\thesection}{#1} }

\catcode`\@=11
\def\numberbysection{\@addtoreset{equation}{section}
        \def\theequation{\thesection.\arabic{equation}}}
\numberbysection

\begin{document}

\title[Phase-separation in the critical spherical model]{Kinetics of 
phase-separation in the critical spherical model and local
scale-invariance}

\author{Florian Baumann$^{a,b}$ and 
Malte Henkel$^a$} 
\address{$^a$Laboratoire de Physique des 
Mat\'eriaux,\footnote{Laboratoire associ\'e au CNRS UMR 7556} 
Universit\'e Henri Poincar\'e Nancy I, \\ 
B.P. 239, F -- 54506 Vand{\oe}uvre l\`es Nancy Cedex, France}
\address{$^b$Institut f\"ur Theoretische Physik I, 
Universit\"at Erlangen-N\"urnberg, \\
Staudtstra{\ss}e 7B3, D -- 91058 Erlangen, Germany}


\begin{abstract}
The scaling forms of the space- and time-dependent two-time correlation and 
response functions are calculated for the kinetic spherical model with a 
conserved order-parameter and quenched to its critical point from a completely 
disordered initial state. The stochastic Langevin equation can be split into
a noise part and into a deterministic part which has local scale-transformations with a dynamical 
exponent $z=4$ as a dynamical symmetry. An exact reduction formula allows to
express any physical average in terms of averages calculable from the 
deterministic part alone. The exact spherical model results are shown to
agree with these predictions of local scale-invariance. The
results also include kinetic growth with mass conservation
as described by the Mullins-Herring equation.
\end{abstract}
\submitto{J. Stat. Mech.}
\maketitle

\setcounter{footnote}{1}


\section{Introduction}

A many-body system rapidly brought out of an initial equilibrium state by
quenching it to either a critical point or into a coexistence  region of the
phase-diagramme where there are at least two equivalent equilibrium states
undergoes {\em ageing}, that is the physical state depends on the time since
the quench was performed and hence time-translation invariance 
is broken. Furthermore,
either the competition between the distinct 
equilibrium states or else non-equilibrium
critical dynamics leads to a slow dynamics, where  observables depend
non-exponentially on time. Both aspects of ageing can be conveniently 
studied through the two-time response and correlation functions defined as
\BEQ
\label{scaling_r}
R(t,s) = \left.\frac{\delta\langle\phi(t,\vec{r})\rangle}{\delta h(s,\vec{r})}
\right|_{h=0} = s^{-a-1} f_R\left(\frac{t}{s}\right) \;\;,\;\;
f_R(y) \stackrel{y \rightarrow \infty}{\sim}
y^{-{\lambda_R}/{z}}
\EEQ
\BEQ
\label{scaling_c}
C(t,s) = \bigl\langle \phi(t,\vec{r})\phi(s,\vec{r})\bigr\rangle 
= s^{-b} f_C\left(\frac{t}{s}\right) \;\;,\;\;
f_C(y) \stackrel{y \rightarrow \infty}{\sim} y^{-{\lambda_C}/{z}}
\EEQ
where  $\phi(t,\vec{r})$ is the order-parameter at time $t$ and location
$\vec{r}$ and $h$ is the conjugate magnetic field. 
These dynamical scaling forms (\ref{scaling_r},\ref{scaling_c}) 
are expected  to hold in the 
scaling limit where both $t,s\gg t_{\rm micro}$ 
and also $t-s\gg t_{\rm micro}$,
where $t_{\rm micro}$ is some microscopic reference time. In writing
those, it is implicitly assumed that the underlying
dynamics is such that there is a single relevant length-scale 
$L=L(t)\sim t^{1/z}$,
where $z$ is the dynamical critical exponent. Universality classes of ageing
are defined through the values of the non-equilibrium exponents 
$a,b,\lambda_C,\lambda_R$. for a non-vanishing initial value of the order-parameter, it is known that $\lambda_{C,R}$ may be related to the 
slip exponent introduced in \cite{Jans89} but there are no scaling
relations with equilibrium exponents. For reviews, see e.g.
\cite{Bray94,Godreche02,Calabrese05,Cugl02,Lux}. 

A lot of attention has been devoted to the case 
where the underlying dynamics is
such that there are no macroscopic conservation laws and in particular the 
order-parameter is non-conserved. In this situation, which is appropriate for
the description of the ageing of magnets, results for the exponents 
$a,b,\lambda_C,\lambda_R$ have been derived. 
Furthermore, it appears that simple
dynamical scaling can be extended to a local scale-invariance, where the
response functions  transform covariantly  under more general time-dependent 
rescalings of time and space \cite{Henkel02}. In particular, this leads to 
explicit predictions for the scaling functions  $f_R(y)$ (and also for 
$f_C(y)$ if $z=2$) which  have been tested and confirmed in a large variety of
systems, see \cite{Henkel06} for a recent review. 

In this paper, we are interested in carrying out a first 
case-study on the ageing
and its local scale-invariance 
when the order-parameter is conserved. Physically, 
this may describe the phase-separation
of binary alloys, where the order-parameter is given by the 
concentration difference
of the two kinds of atoms. Another example are kinetic
growth-processes at surfaces when mass conservation holds.
One of the most simple models of this kind is the
Mullins-Herring model \cite{Mul63,Wol90}, which leads to a
dynamical exponent $z = 4$ and has recently been discussed
in detail in \cite{Roethlein06}.

Presently, there exist very 
few studies on ageing with
a conserved order-parameter. In a numerical study of the $2D$ Ising model 
quenched to $T<T_c$, where $z=3$, 
Godr\`eche, Krzakala and Ricci-Tersenghi \cite{Godreche04} 
calculated  the scaling function $f_C(y)$ which remarkably shows a cross-over
between a first power-law decay 
$f_C(y) \sim y^{-\lambda_C'/z}$ for intermediate
values of $y$ to the final asymptotic behaviour
(\ref{scaling_c}) for $y\gg 1$. They find a value of
$\lambda_C' = 2.5$, whereas the value for $\lambda_C$ is
much smaller with $\lambda_C = 1$. 
On the other hand, Sire \cite{Sire04} calculated $f_C(y)$ exactly in the
critical spherical model ($T=T_c$), where $z=4$, and found a single power-law 
regime.\footnote{It is well-known that the spherical model with conserved
order-parameter quenched to below $T_c$ shows a multiscaling which is not
captured by the simple scaling  form (\ref{scaling_c}) \cite{Coniglio89}. It
is understood that this is a peculiarity of the $n\to\infty$ limit of the
conserved O($n$) vector-model \cite{Mazenko90}.} For a review, see
\cite{Calabrese05}.

Here, we shall revisit the
spherical model with a conserved order-parameter, quenched
to $T=T_c$. Besides reviewing the calculation of the
two-time correlation function we shall
also provide the exact solution for the two-time response function. This will be
described in section 2. Our main focus will be to show how these  exact results
can be understood from an extension of dynamical scaling to 
local scale-invariance (LSI) with
$z=4$. In section 3, we recall the main results of 
LSI for that case and in particular
write down a linear partial differential equation of which this LSI is a
dynamical symmetry. In section 4, we consider the 
Langevin equation which describes
the kinetics with a conserved order-parameter (model B in the Halperin-Hohenberg
classification) and show for the case at hand how the
Langevin-equation can be split into a `deterministic' and a
`noise' part, so that all averages to be calculated can be exactly reduced
to quantities to be found in a noise-less  theory. 
Local scale-invariance directly
determines the latter, as we shall see in section 5, 
where we can finally compare
with the exact spherical model results of section 2. This
is, together with the earlier study of the Mullins-Herring model
in \cite{Roethlein06}, the first 
example where local scale-invariance can be confirmed for
an exactly solvable model with dynamical exponent $z\ne 2$. 
Our conclusions are formulated in
section 6. Some technical points are reported in the
appendices. In appendix A we derive the two-point function
from the symmetry operators and in appendix B we calculate
solution of an integral which appears frequently in our
computations. In an outlook beyond the present specific
context, we discuss in appendix C the
autoresponse function for values of $z$ different from $4$.
In appendix D we discuss the most simple possible model,
namely a free particle submitted to conserved noise. This
serves as a further illustration of some of the space-time
properties of the conserved spherical model.

\section{The spherical model with a conserved order-parameter}

The spherical model was conceived in 1953 by Berlin and Kac as a mathematical
model for strongly interacting spins which is yet easily solvable and it has
indeed served a useful r\^ole in providing exact results in a large variety
of interesting physical situations. It may  be defined in terms of a real
spin variable $S(t,\vec{x})$ attached to each site $\vec{x}$ of a 
hypercubic lattice $\Lambda \subset \mathbb{Z}^d$ and depending on 
time $t$, subject to the (mean) spherical constraint
\BEQ
\left\langle \sum_{\vec{x}\in\Lambda} S(t,\vec{x})^2 \right\rangle 
= {\cal N}
\EEQ
where $\cal N$ is the number of sites of the lattice. The Hamiltonian
is ${\cal H} = -\sum_{(\vec{x},\vec{y})} S_{\vec{x}} S_{\vec{y}}$ where
the sum is over pairs of nearest neighbours. 
The kinetics is assumed to be given by a Langevin equation of model B type 
\BEQ
\label{spherical}
\partial_t S(t,\vec{x}) = -
\nabla_{\vec{x}}^2 \left[\nabla_{\vec{x}}^2 S(t,\vec{x}) +
\mathfrak{z}(t) S(t,\vec{x}) + h(t,\vec{x})\right] + \eta(t,\vec{x})
\EEQ
where $\mathfrak{z}(t)$ is the Lagrange multiplier fixed by the mean spherical 
constraint\footnote{Considering the spherical constraint in the mean simplifies 
the calculations and should not affect the scaling behaviour. See 
\cite{Fusco02} for a careful study of this point in the non-conserved case.} 
and the coupling to the heat bath with the {\it critical} temperature $T_c$ is 
described by a Gaussian noise $\eta$ of vanishing average and variance
\BEQ
\label{noise_corr}
\left\langle \eta(t,\vec{x})\eta(t',\vec{x}')\right\rangle = 
-2T_c \nabla_{\vec{x}}^2\delta(t-t')
\delta(\vec{x}-\vec{x}'). 
\EEQ
and $h(t,\vec{x})$ is a small external magnetic field.\footnote{Since we are
only interested in averages of local quantities and the initial magnetisation 
is assumed to vanish, this description is sufficient and we need not consider
the analogue of the more elaborate equations studied by Annibale and Sollich 
\cite{Annibale06} for the non-conserved case.}
Here the derivative on the right-hand side of
(\ref{noise_corr}) expresses the
fact that the noise is chosen in such a way that it does not break
the conservation law.  
On the other hand, there are physical situations such as
surface growth with mass conservation, where this is not the
case. The non-conserved Mullins-Herring model
\cite{Mul63,Wol90} is given by equation
(\ref{spherical}) with $\mathfrak{z} = 0$ and the noise correlator $\langle
\eta(t,\vec{x}) \eta(t',\vec{x}') \rangle = 2 T_c \, \delta(t-t')
\delta(\vec{x}-\vec{x}')$, see \cite{Roethlein06} for
details. Here we shall study the conserved Mullins-Herring model, that is
equation (\ref{spherical}) with $\mathfrak{z}(t) = 0$ and the noise
correlator (\ref{noise_corr}).
A similar remark applies on the way we included the perturbation
$h(t,\vec{x})$. Here we study the case, where the
perturbation does not break the conservation law.

The spherical constraint is taken into account through the
Lagrange multiplier $\mathfrak{z}(t)$ which has to be
computed self-consistently. 
These equations have already been considered several times in the literature
\cite{Kissner92,Majumdar95,Sire04}. 

\subsection{The correlation function}

We repeat here the main steps of the calculation of the correlation 
function \cite{Kissner92,Majumdar95,Sire04}. 
Equation (\ref{spherical}) with $h(t,\vec{x}) = 0$ is readily solved in
Fourier-space, yielding 
\BEQ
\label{solution_spherical}
\wht{S}(t,\vec{k}) = \left[\wht{S}(0,\vec{k}) + \int_0^t \!\D\tau\:
\exp \Big(\omega(\tau,\vec{k})\Big) \eta(\tau,\vec{k})
\right]\exp\Big(-\omega(t,\vec{k})\Big)
\EEQ
where $\wht{S}(t,\vec{k})$ denotes the spatial Fourier transformation of
$S(t,\vec{r})$ and 
\BEQ
\omega(t,\vec{k}) = k^4 t - k^2 \int_0^t \!\D \tau\,
\mathfrak{z}(\tau), \qquad \mbox{with} \qquad k := |\vec{k}|
\EEQ
Next, $\omega(t,\vec{k})$ is determined from the spherical constraint. This
is easiest in the continuum limit where spatial translation-invariance
leads to $\langle S^2(t,\vec{x})\rangle = (2\pi)^{-d}\int^\Lambda \!\D
\vec{k}\,\langle \wht{S}(t,-\vec{k}) \wht{S}(t,\vec{k}) \rangle=1$ 
where $\Lambda$ is the inverse of a lattice cutoff. This has already been 
analysed in the literature \cite{Kissner92,Majumdar95,Sire04}, with the result
\BEQ
\omega(t,\vec{k}) = k^4 t - \mathfrak{g}_d k^2
t^{{1}/{2}}
\EEQ
where the constant $\mathfrak{g}_d$ is determined by the condition \cite{Kissner92} 
\BEQ
\hspace{-1.2truecm} 
\int_0^{\infty} \!\D x\, x^{d-1} \left\{ 2x^2 \int_0^1 \!\D y\, 
\exp\left[ -2x^4 (1-y) + 2 \mathfrak{g}_d x^4 (1- \sqrt{y\,}) \right] 
-\frac{1}{x^2} \right\} = 0
\EEQ
In particular, $\mathfrak{g}_d=0$ for $d > 4$ and asymptotic forms for $d\to 2$ and 
$d\to 4$ are listed in \cite{Kissner92}.  
Then from
(\ref{solution_spherical}) the two-time correlator
$\wht{C}(t,s,\vec{k}) := \Big\langle \wht{S}(t,-\vec{k}) \wht{S}(s,\vec{k})
\Big\rangle$ is straightforwardly derived 
\BEA
\label{result_A}
\lefteqn{ \wht{C}(t,s,\vec{k}) = s_0(\vec{k})^2
\exp\Big(-\omega(t,\vec{k})-\omega(s,\vec{k})\Big) 
}
\nonumber
\\ &+& 2 T_c \int_0^s \!\D u\:
k^2 \exp\Big(-\omega(t,\vec{k}) - \omega(s,\vec{k}) + 2
\omega(u,\vec{k})\Big)
\EEA
where $s_0(\vec{k})^2 := 
\Big\langle \wht{S}(0,-\vec{k})\wht{S}(0,\vec{k})\Big\rangle$.
Transforming back to direct space, one obtains
\BEQ
C(t,s;\vec{r}) = A_1(t,s;\vec{r}) + A_2(t,s;\vec{r})
\EEQ
where $A_1(t,s;\vec{r})$ and $A_2(t,s;\vec{r})$ are the
inverse
Fourier transforms of the first and the second line of the
right-hand side of equation (\ref{result_A}), respectively.
They are explicitly given by the following expressions,
where we use the shorthand $y := t/s$
\BEA
\label{a1}
A_1(t,s;\vec{r})& =&  (t+s)^{-{d}/{4}} 
\int_{\mathbb{R}^d} \frac{\D \vec{k}} {(2 \pi)^d }\, s_0(\vec{k})^2
\exp \left(-\frac{\II \vec{k}\cdot\vec{r}}{(t+s)^{{1}/{4}}}\right) 
\\ \nonumber
& &\times \exp\left(-k^4 + \mathfrak{g}_d k^2 \frac{y^{{1}/{2}}
+1}{(y+1)^{{1}/{2}}} \right)
\EEA
and
\BEA
\label{a2}
A_2(t,s;\vec{r}) &=& 2 T_c s^{-(d-2)/4}  \int_0^1
\!\D \theta \; (y+1-2 \theta)^{-(d+2)/4}
\int_{\mathbb{R}^d} \frac{\D \vec{k}} {(2 \pi)^d }\, k^2 \\ & &
\hspace{-1.5cm} \times \exp\left( -\frac{\II \vec{k}\cdot \vec{r}}{(s (y+1 - 2
\theta))^{{1}/{4}}} \right) 
\exp \left(-k^4 + \mathfrak{g}_d k^2 
\left(\frac{ y^{1/2} +1 - 2\theta^{1/2} }{(y+1-2 \theta)^{{1}/{2}}} \right)
\right) \nonumber
\EEA
The term $A_2(t,s;\vec{r})$ is the dominating one in the scaling
regime where $t,s\to\infty$ and $y=t/s$ is kept fixed. 
There we recover the scaling form (\ref{scaling_c}) with $b = (d-2)/4$. 
The explicit expression for the scaling function
$f_C(y)$ is quite cumbersome but, if the last term in the second exponential
in eq.~(\ref{a2}) can be treated as a constant, one has, up to normalisation  
\BEQ
\label{slf_fc}
f_C(y) \sim \left\{
\begin{array}{cc}
\frac{8 T_c}{2-d} \left[ (y-1)^{(2-d)/4}  -
(y+1)^{(2-d)/4}  \right] & \qquad \mbox{\rm ;~~ $d > 2$} \\
2 T_c \ln\left(\frac{y-1}{y+1}\right) & \qquad \mbox{\rm ;~~ $d = 2$}
\end{array}
\right.
\EEQ
This is exact for the spherical model for $d>4$, since then $\mathfrak{g}_d=0$. 
Equation (\ref{slf_fc}) also applies to the Mullins-Herring
equation for any $d \geq 1$ (since here $\mathfrak{z} = 0$, in
this case $\mathfrak{g}_d = 0$ exactly). For $y$ sufficiently large,
the above condition is also approximately satisfied. In any case, in the limit
$y\to\infty$ we can read off the autocorrelation exponent 
$\lambda_C = d +2$. 

\begin{figure}[htb] 
  \vspace{0.5cm}
  \centerline{\epsfxsize=6.0in\epsfclipon\epsfbox
  {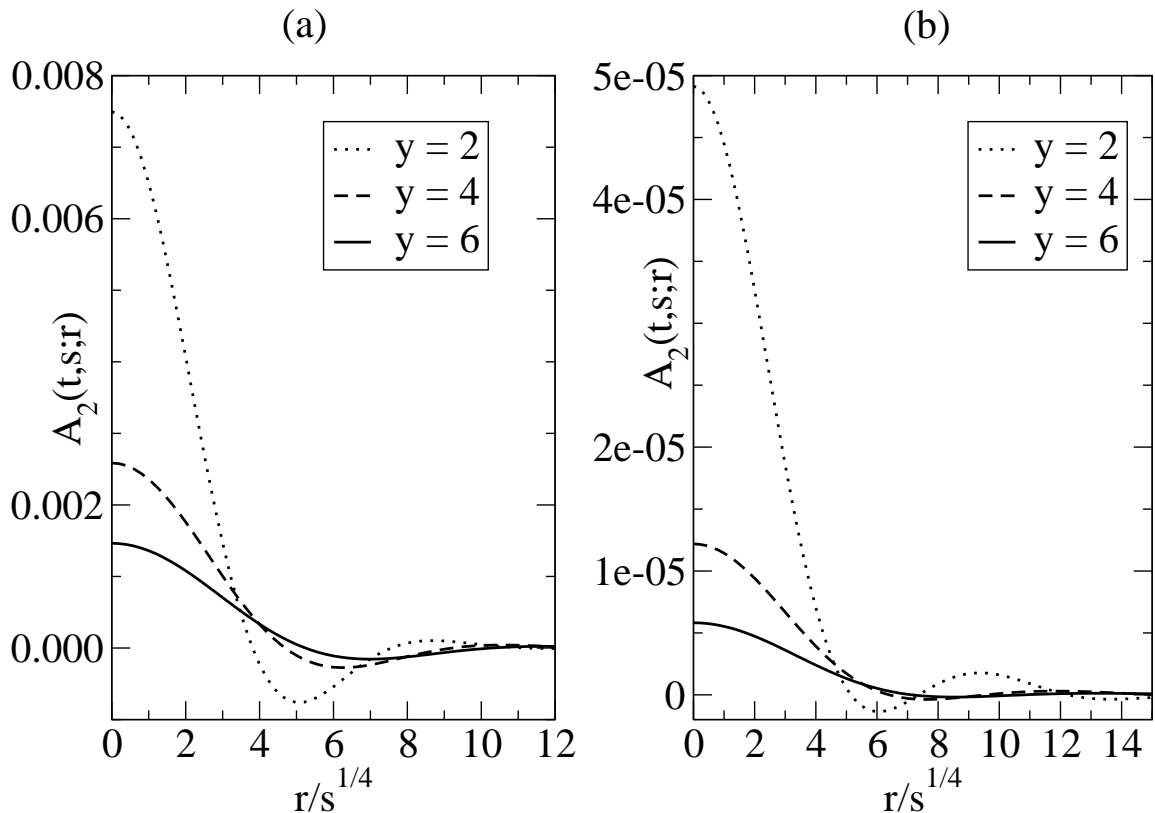}
  }
  \caption[figure]{Scaling of the spatio-temporal correlation function
  $C(t,s;\vec{r})\simeq A_2(t,s;\vec{r})$ with $s=20$ and for several values 
  of $y=t/s$ for (a) $d=3$ (left panel) and (b) $d=5$ (right panel).   
  \label{figure2}}
\end{figure} 

The scaling behaviour of the spatio-temporal two-time correlator is illustrated
in figure~\ref{figure2}. We see that for relatively small values of $y=t/s$, the
correlation function does not decay monotonously towards zero, but rather
displays oscillations whose amplitude decays with $|\vec{r}| s^{-1/4}$. This 
behaviour is quite close to the well-known one for the equal-time correlation
function. As $y$ increases, these oscillations become less pronounced. There
is no apparent qualitative difference between the cases $d<4$ and $d>4$.
 
\subsection{The response function}

The response function can be computed in a way similar to the
non-conserved case \cite{Godreche00}. Equation (\ref{spherical}) is solved
in Fourier space yielding an equation similar to (\ref{solution_spherical})
\BEA
\wht{S}(t,\vec{k}) &=& \Big[\wht{S}(0,\vec{k}) + \int_0^t \!\D\tau\,
		   \exp\Big(\omega(\tau,\vec{k})\Big)
		   \nonumber \\ && \times  
		   \Big(\wht{\eta}(\tau,\vec{k})+
		   k^2\wht{h}(\tau,\vec{k})\Big)
                   \Big]\exp(-\omega(t,\vec{k}))
\EEA
{}From this the Fourier transform of the response function is computed as 
$\wht{R}(t,s,\vec{k}) = {\delta \langle \wht{S}(t,\vec{k})
\rangle}/{\delta h(s,-\vec{k})}$. Transforming the result back to direct space,
we obtain for $t > s$
\BEA
\label{response_result}
R(t,s;\vec{r}) &=& (t-s)^{-(d+2)/4} \int_{\mathbb{R}^d}
\frac{\D\vec{k}}{(2\pi)^d}k^2 \exp \left(-\frac{\II \vec{k} \cdot
\vec{r}}{(t-s)^{{1}/{4}}}\right) \nonumber \\
& & \times \exp\left(-k^4 +
\mathfrak{g}_d k^2 \frac{y^{{1}/{2}}-1}{(y-1)^{{1}/{2}}}\right)
\EEA
It is easy to see that the scaling form
(\ref{scaling_c}) holds with $ a = (d-2)/4 =b$. 
For large values of $y$ the scaling function $f_R(y)$ is
given by
\BEQ
f_R(y) \sim (y-1)^{-{(d+2)}/{4}}.
\EEQ
Again, for $d > 4$ this expression is exact up to a
prefactor. The value of the autoresponse exponent therefore is, for $d>2$  
\BEQ
\lambda_R = d + 2 = \lambda_C
\EEQ
in agreement with field-theoretical expectations for the O($n$) model
\cite{Calabrese05}. 


\begin{figure}[htb] 
  \vspace{0.5cm}
  \centerline{\epsfxsize=6.0in\epsfclipon\epsfbox
  {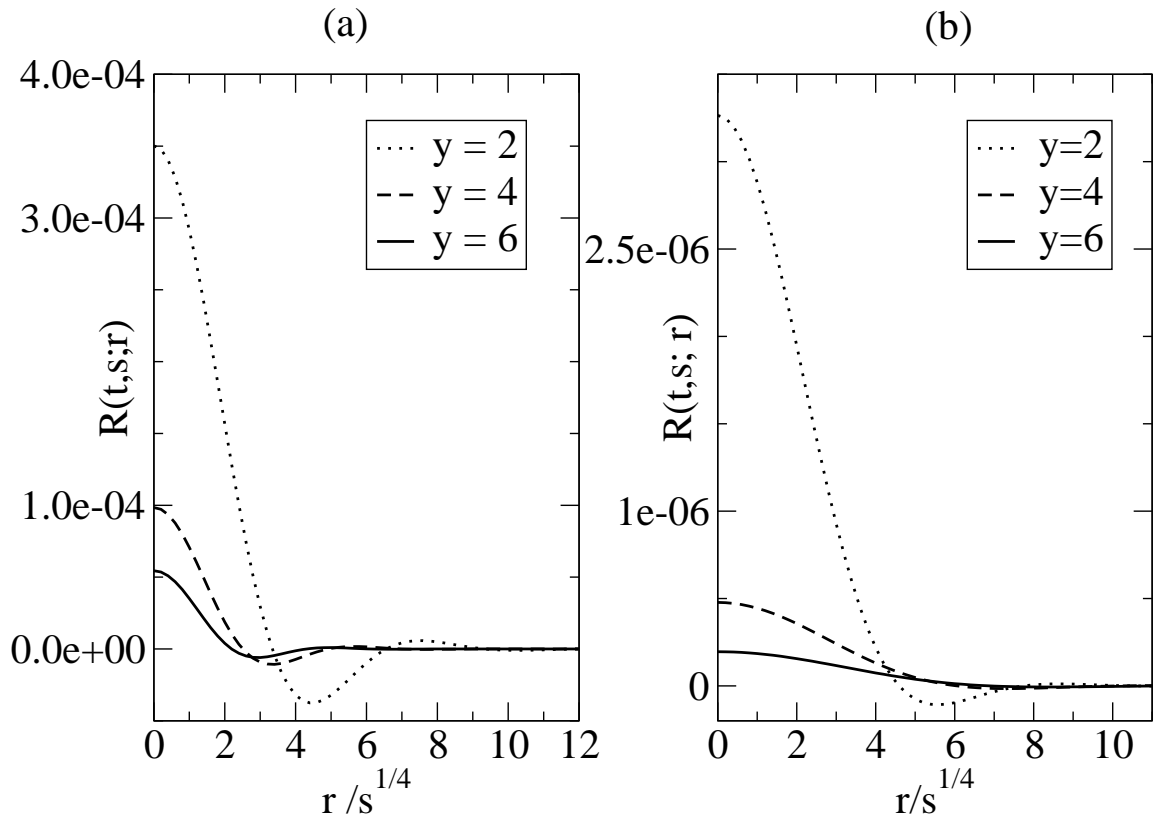}
  }
  \caption[figure]{Scaling of the spatio-temporal response function
  $R(t,s;\vec{r})$ for $s=20$ and several values of $y=t/s$ for $d=3$ 
  (left panel) and $d=5$ (right panel).
  \label{figure1}}
\end{figure} 

It is instructive to consider the full space-time
response function explicitly for the case that $\mathfrak{g}_d
= 0$, that is for $ d > 4$ in the spherical model and for
any $d$ in the Mullins-Herring equation. The integral in
(\ref{response_result}) is done in appendix~B and we find
\BEA
\label{result_response}
\lefteqn{ R(t,s;\vec{r}) = 
\frac{\sqrt{\pi}
}{2^{3d/2} \pi^{d/2} \Gamma(d/4)}\: (t-s)^{-{(d+2)}/{4}}
\left[{_0F_2}\left(
\frac{1}{2},\frac{d}{4};\frac{r^4}{256\,
(t-s)}\right)\right.  } 
\nonumber \\
&-& \left.\frac{8}{d}
\frac{\Gamma\left(\frac{d}{4}+1\right)}{\Gamma
\left(\frac{d}{4}+\frac{1}{2}\right)}
\left(\frac{r^2}{16 \sqrt{t-s}} \right)
{_0F_2}\left(\frac{3}{2},\frac{d}{4}+\frac{1}{2};
\frac{r^4}{256 \, (t-s)}\right)\right]
\EEA
This expression will be useful later for direct comparison with the 
results obtained from the theory of
local scale-invariance. For $\mathfrak{g}_d \neq 0$ it is more
convenient to use the integral representation
(\ref{response_result}). In figure~\ref{figure1} we display the spatio-temporal
response function. In contrast to the non-conserved case for a fixed value
of $y=t/s$ there are decaying oscillations with the scaling variable
$|\vec{r}| s^{-1/4}$. These oscillations disappear rapidly when $y$ is
increased. This qualitative behaviour also arises, as we
show in appendix D, for the simple random walk with a
conserved noise. 

Having found both correlation and response functions, we can also
calculate the fluctuation-dissipation ratio $X(t,s) = T R(t,s) (\partial
C(t,s)/\partial s)^{-1}$ which measures the distance from equilibrium
\cite{Cugl94}. In particular, we find for the fluctuation-dissipation limit
\BEQ
X_{\infty} = \lim_{s\to\infty} \left( \lim_{t\to\infty} X(t,s) \right) = 
\frac{1}{2}
\EEQ
in agreement with the known result \cite{Calabrese05} of the conserved
$n\to\infty$ limit of the O($n$) model.   

\section{Symmetries of the deterministic equation}

Before we can examine dynamical symmetries in Langevin equations, we need
some background material on the dynamical symmetries of the deterministic part 
of such equations. Therefore, we recall in this section first some facts about 
Schr{\"o}dinger-invariance which describes an extension of dynamical scaling
for a dynamical exponent $z=2$. As shown earlier \cite{Henkel02}, an analogous 
extension can be built for $z\ne 2$ and we shall write down the relevant
generators for this {\em local scale-invariance} for $z=4$. 

The physical starting point is the covariance of $n$-point functions such as
the correlator $F^{(n)}(t_1,\ldots,t_n;\vec{r}_1,\ldots,\vec{r}_n,) 
= \langle \phi_1(t_1,\vec{r}_1) \ldots,\phi_n(t_n,\vec{r}_n) \rangle$
under the rescaling of spatial and temporal coordinates 
\BEQ
\hspace{-1.0cm}
F^{(n)}(b^z t_1,\ldots , b^z t_n; b \vec{r}_1, \ldots, b \vec{r}_n) =
b^{-(x_1+\ldots+x_n)} F^{(n)}(t_1,\ldots t_n;\vec{r}_1, \ldots ,
\vec{r}_n)
\EEQ
where the $x_i$ are the scaling dimensions of the fields $\phi_i$ and $z$
is the dynamical exponent. By analogy with conformal invariance {\em at}
equilibrium, one asks whether an extension of dynamical scaling to 
local scale-transformations with a space-time dependent rescaling factor 
$b(t,\vec{r})$ might be possible. The most simple case is the one of
{\em Schr\"odinger-invariance} which describes the dynamical symmetries of the
linear diffusion equation (or free Schr\"odinger equation)
\BEQ
\label{schroedinger1}
\hat{S}_{\rm sch} \phi(\vec{r},t) = 0,\qquad \mbox{with} \quad 
\hat{S}_{\rm sch} := \lambda \partial_t - \frac{1}{4}
\nabla_{\vec{r}}^2
\EEQ
where the parameter $\lambda$ is related to the inverse diffusion constant.
The symmetry group of this equation, that is the largest group of
transformations carrying solutions of (\ref{schroedinger1})
to other solutions, was already found 1882 by Lie and is the well-known 
Schr{\"o}dinger group {\sl Sch}($d$), see \cite{Niederer72}.
It transforms coordinates as $(t,\vec{r}) \mapsto
(t',\vec{r}') = g(t,\vec{r})$ with
\BEQ
\vec{r}' = \frac{\mathcal{R} \vec{r} + \vec{v} t +
\vec{a}}{\tilde{\gamma} t + \tilde{\delta}} \;\; , \;\; 
t' = \frac{\tilde{\alpha} t +
\tilde{\beta}}{\tilde{\gamma} t + \tilde{\delta}} \;\; ; \;\;
\tilde{\alpha} \tilde{\delta} - \tilde{\beta} \tilde{\gamma} = 1
\EEQ
where $\mathcal{R}$ is a rotation matrix and $\vec{v},
\vec{a},\tilde{\alpha},\tilde{\beta},\tilde{\gamma}$ 
and $\tilde{\delta}$ are parameters. Evidently, the dynamical
exponent is $z=2$. The solutions of (\ref{schroedinger1}) transform as
\BEQ
\label{quasiprimary1}
\phi(t,\vec{r}) \rightarrow (T_g \phi)(t,\vec{r}) =
f_g[g^{-1}(t,\vec{r})] \phi(g^{-1}(t,\vec{r}))
\EEQ
where the companion function $f_g$ is known explicitly
\cite{Niederer72}.
A field transforming as (\ref{quasiprimary1}) is called {\it quasiprimary}. 
The generators of the Schr{\"o}dinger group form a Lie
algebra $\mathfrak{sch}_d= \mbox{\rm Lie({\sl Sch}($d$))}$ 
and it can be shown that a $n$-point correlator 
built from quasiprimary fields must satisfy the conditions
\BEQ
\label{quasiprimary2}
\sum_{i=1}^n \mathcal{X}_i \; \left\langle \phi_1(t_1,\vec{r}_1) \ldots
\phi_n(t_n,\vec{r}_n) \right\rangle = 0
\EEQ
where $\mathcal{X}_i$ stands for any generator of $\mathfrak{sch}_d$
which acts on the field $\phi_i$. For example, eq.~(\ref{quasiprimary2}) 
fixes the two-point function of quasiprimary fields completely \cite{Henkel94}. 

Following \cite{Henkel02}, we now write the generalisation of this to
the case when $z=4$ which has already been started in \cite{Roethlein06}. 
Consider the `Schr\"odinger operator'
\BEQ
\label{diff_operator}
\hat{S} := -\lambda \partial_t + \frac{1}{16}
\left(\nabla_{\vec{r}}^2\right)^2.
\EEQ
which we shall encounter in the context of the conserved spherical model
and where obviously $z=4$. 
Using the shorthands $\vec{r} \cdot \partial_{\vec{r}} := \sum_{k=1}^d r_k\,
\partial_{r_k}$,$\nabla_{\vec{r}}^2 := \sum_{k=1}^d\partial_{r_k}^2 $ and 
$\vec{r}^2 := \sum_{k=1}^d r_k^2$, the infinitesimal generators of local
scale-transformations read (using a notation analogous to \cite{Henkel02}) 
\BEA
X_{-1} & := & -\partial_t  \nonumber \\
X_0 & := & -t \partial_t - \frac{1}{4}\, \vec{r} \cdot
\partial_{\vec{r}} - \frac{x}{4} \nonumber \\
X_1 & := & - t^2 \partial_t - \frac{
x}{2} t - \lambda \vec{r}^2 \,
(\nabla_{\vec{r}}^2)^{-1}
- \frac{1}{2} \, t \vec{r} \cdot \partial_{\vec{r}}
\nonumber \\ 
& & +4 \gamma \, \left( \vec{r}\cdot \partial_{\vec{r}}
\right) \, \left( \nabla_{\vec{r}}^2 \right)^{-2}  +
2\gamma (d-4) \left( \nabla_{\vec{r}}^2 \right)^{-2}
\label{gl:alg4} \\
R^{(i,j)} &:=& r_i \partial_{r_j} - r_j \partial_{r_i}
\;\; ; \;\; \mbox{\rm where $1\leq i < j\leq d$} 
\nonumber \\
Y^{(i)}_{-1/4} & = & - \partial_{r_i} \nonumber \\
Y^{(i)}_{3/4} & = & - t \partial_{r_i} - 4 \lambda
r_i \left( \nabla_{\vec{r}}^2
\right)^{-1} + 8 \gamma \, 
\partial_{r_i} \left( \nabla_{\vec{r}}^2 \right)^{-2} \nonumber
\EEA
where $x$ is the scaling dimension of the fields on which these
generators act and $\gamma,\lambda$ are further field-dependent parameters.
Here, the generators $X_{\pm 1,0}$ correspond to projective changes in 
the time $t$, the generators $Y^{(i)}_{n-1/4}$ are space-translations, 
generalised Galilei-transformations and so on 
and $R^{(i,j)}$ are spatial rotations. 
In writing these generators, the following properties of the 
derivative $\partial_r^{\alpha}$ are assumed
\BEQ
\partial_r^{\alpha} \partial_r^{\beta} = \partial_r^{\alpha+\beta} 
\;\; , \;\;
{} \left[ \partial_r^{\alpha}, r \right] = \alpha \partial_r^{\alpha-1}
\EEQ
and which can be justified in terms of fractional derivatives as shown
in detail in appendix~A of \cite{Henkel02}. The operator 
$(\nabla_{\vec{r}}^2)^{-1}$ can then be defined formally as follows.
For example, in $d=2$ dimensions, we have
\BEQ
(\nabla_{\vec{r}}^2)^{-1} := (\partial_{r_x}^2 +
\partial_{r_y}^2)^{-1} := \sum_{n=0}^\infty (-1)^n
\partial_{r_x}^{-2-2n} \partial_{r_y}^{2 n}
\EEQ
The remaining negative powers of $\nabla_{\vec{r}}$ are then defined by 
concatenation, e.g. $(\nabla_{\vec{r}}^2)^{-2} =
(\nabla_{\vec{r}}^2)^{-1} \cdot (\nabla_{\vec{r}}^2)^{-1}$.
We easily verify the following commutation relations for $n\in\mathbb{Z}$
\BEA
\label{commutators1}
{[}(\nabla_{\vec{r}}^2)^n,r_i] &=& n\, \partial_{r_i}
(\nabla_{\vec{r}}^2)^{n-1} \\
\label{commutators2}
{[}(\nabla_{\vec{r}}^2)^n,\vec{r}^2] & = & 2 n \, (\vec{r} \cdot
\partial_{\vec{r}}) \, (\nabla_{\vec{r}}^2)^{n-1} + n \, (n-2+d)
(\nabla_{\vec{r}}^2)^{n-1}
\EEA

The generators eq.~(\ref{gl:alg4}) describe dynamical symmetries of the 
`Schr\"odinger operator' (\ref{diff_operator}). This can be seen from the
commutators of $\hat{S}$ and the generators ${\cal X}$ 
from eq.~(\ref{gl:alg4}). Straightforward, but a little tedious
calculations give, analogously to \cite{Henkel02}
\BEA
{} [\hat{S},X_{-1}] &=& 0 \;\; , \;\; [\hat{S},Y^{(i)}_{-1/4}] \:=\: 0  
   \;\; , \;\; [\hat{S},Y^{(i)}_{3/4}] \:=\: 0 \nonumber \\
{} [\hat{S},X_0] &=& - \hat{S} \;\; , \;\; [\hat{S},R^{(i,j)}] \:=\: 0 
\EEA
This means that for a solution of the `Schr\"odinger equation' $\hat{S}\phi=0$ 
the transformed function ${\cal X}\phi$ is again solution of
the `Schr\"odinger equation'. 
For the commutator with $X_1$ we find 
\BEQ
{}   [\hat{S},X_1] = -2 t \hat{S} + \frac{\lambda}{2} 
   \left( x - \frac{d}{2} -1 +\frac{2\gamma}{\lambda} \right) 
\EEQ
hence a dynamical symmetry is found if the field $\phi$ has the scaling
dimension
\BEQ
\label{rel_scaling_dim}
x = \frac{d}{2} + 1-\frac{2 \gamma}{\lambda}
\EEQ
which for $d=1$ reproduces the result of \cite{Henkel02}.\footnote{For
a free field-theory, where $x=d/2$, this implies $\gamma/\lambda=\frac{1}{2}$.} 
Generalising from conformal or Schr\"odinger-invariance, quasiprimary
fields transform covariantly and their $n$-point functions will
again satisfy eq.~(\ref{quasiprimary2}). A quasiprimary field
is now characterised by its scaling dimension $x_i$ and the further
parameters $\gamma_i,\lambda_i$. For example, any two-point function
$F^{(2)} = \langle \phi_1 \phi_2\rangle$ built from two quasiprimary fields
$\phi_{1,2}$ is completely fixed by solving the 
conditions (\ref{quasiprimary2}) for the generators in (\ref{gl:alg4}). 
We quote the
result and refer to appendix~A for the details of the
calculation.
\BEA
\lefteqn{ 
F^{(2)}(t-s,\vec{x}-\vec{y}) := \langle \phi_1(t,\vec{x})
\phi_2(s,\vec{y})\rangle 
}
\nonumber \\
&=& \delta_{x_1,x_2}\delta_{\lambda_1,-\lambda_2}
\delta_{\gamma_1,-\gamma_2}\,
(t-s)^{-{x}/{2}} \sum_{s \in \mathcal{E}'} c_s 
\phi^{(s)}\left( |\vec{x}-\vec{y}| (t-s)^{-1/4}\right)
\EEA
Here $\phi^{(s)}(u)$ are scaling functions, the $c_s$ are free parameters 
and the set $\mathcal{E}'$ is defined as follows
\BEQ
   \mathcal{E}' := \left\{ \begin{array}{lcl} \{2,4\} &
                                \mbox{if} & d > 4 \\
                            \{2,4,4-d\} & \mbox{if} & 2 <
			    d \leq 4\\
                            \{2,4,2-d,4-d\}& \mbox{if} & d
			    \leq 2 
                           \end{array} \right.
\EEQ
We shall see later that boundary conditions may impose further conditions on 
the $c_s$. The functions $\phi^{(s)}(u)$ are given by the series, convergent
for all $|u|<\infty$
\BEQ
\phi^{(s)}(u) = \sum_{\ell = 0}^\infty b_{\ell}^{(s)} u^{4 \ell + s-4}.
\EEQ
with the coefficients $b_{\ell}^{(s)}$ 
\BEQ
\hspace{-1.0cm}
b_{\ell}^{(s)} =  2^4 
\frac{ \Gamma(\frac{s}{2}+1)
\Gamma(\frac{s}{2}+\frac{d}{2}) }
{\Gamma(\frac{s+d}{4} - \frac{1}{2} - \frac{\gamma}{\lambda})}
\cdot
\frac{\Gamma(\ell +\frac{s+d}{4} - \frac{1}{2} 
-\frac{\gamma}{\lambda})}{\Gamma(2 \ell + \frac{s}{2}-1)
\Gamma(2 \ell + \frac{s}{2} + \frac{d}{2}-2)}\left(-\lambda \right)^{\ell}
\EEQ

\section{Local scale-invariance}

\subsection{General remarks}
Consider the following stochastic Langevin equation
\BEQ
\label{start_equation}
\partial_t \phi = -\frac{1}{16 \lambda} 
\nabla_{\vec{r}}^2 \Big(-\nabla_{\vec{r}}^2 \phi + v(t)
\phi \Big) + \eta
\EEQ
where the noise correlator respects the global conservation law
\BEQ
\langle \eta(\vec{r},t) \eta(\vec{r}',t') \rangle =
-\frac{T}{8 \lambda}
\nabla_{\vec{r}}^2 \delta(\vec{r}-\vec{r}') \delta(t-t')
\EEQ
Here and in what follows, we shall often suppress the
arguments of the fields for the sake of simplicity, if no
ambiguity arises.
We shall adopt the standard field-theoretical setup for
the description of Langevin equations, see e.g.
\cite{Taeuber05,Taeuber07,Taeuber}
for introductions. The Janssen- de Domincis action can be written in 
terms of the order-parameter field $\phi$ and its conjugate response field
$\wit{\phi}$ and reads 
\begin{eqnarray}
\label{action}
\mathcal{J}[\phi,\wit{\phi}] & =& \int \!\D u \D \vec{R}\, \left[
\wit{\phi} \left( \partial_u - \frac{1}{16 \lambda}
\nabla_{\vec{R}}^2 \Big(\nabla_{\vec{R}}^2-v(u)\Big) \right) \phi\right] \\
&+&   \frac{T}{16 \lambda} \int \!\D u \D \vec{R}\:
\wit{\phi}(u,\vec{R}) \,
(\nabla_{\vec{R}}^2 \wit{\phi})(u,\vec{R}) \nonumber
\end{eqnarray}
to which an extra term describing the initial noise  must be added, by analogy with the non-conserved case  \cite{Mazenko98,Picone04}\footnote{It has been 
shown by Janssen that at the initial time $t=0$, the order-parameter field 
$\phi(0,\vec{r})$ and the response field $\wit{\phi}(0,\vec{r})$ are proportional \cite{Janssen92}.} 
\BEQ
\mathcal{J}_{init}[\phi,\wit{\phi}] = \frac{1}{2} \int \!\D\vec{R}
\D\vec{R}'\, \wit{\phi}(0,\vec{R}) \Big\langle \phi(0,\vec{R})
\phi(0,\vec{R}') \Big\rangle \wit{\phi}(0,\vec{R}')
\EEQ
Averages of an observable $\mathcal{O}$ are 
defined as usual by functional integrals with weight
$\exp(-\mathcal{J}[\phi,\wit{\phi}])$, viz. 
\BEQ
\langle \mathcal{O} \rangle := \int \mathcal{D}[\phi] \mathcal{D}[\wit{\phi}]
\mathcal{O} \exp(-\mathcal{J}[\phi,\wit{\phi}])
\EEQ
We decompose the action, in the same way as done in
\cite{Picone04} for the non-conserved case, 
into a deterministic and a noise part, that is
\label{splitup}
\begin{equation}
 \mathcal{J}[\phi,\wit{\phi}] = \mathcal{J}_0[\phi,\wit{\phi}] +
 \mathcal{J}_{\rm b}[\wit{\phi}] 
\end{equation}
with 
\begin{equation}
\mathcal{J}_0[\phi,\wit{\phi}] =
\int \!\D u \D \vec{R}\, \left[
\wit{\phi} \left( \partial_u - \frac{1}{16 \lambda}
\nabla_{\vec{R}}^2 \Big(\nabla_{\vec{R}}^2-v(u)\Big) \right) \phi\right] 
\end{equation}
and ${\cal J}_{\rm b} = {\cal J}_{\rm th} + {\cal J}_{\rm init}$ where 
\begin{equation}
\mathcal{J}_{\rm th}[\wit{\phi}]  =
\frac{T}{16 \lambda} \int \!\D u \D \mathbf{R}\:
\wit{\phi}(u,\mathbf{R}) 
\Big(\nabla_{\vec{R}}^2 \wit{\phi}\Big)(u,\mathbf{R})
\end{equation}
The point of this split-up is, as we shall show, that 
the action ${\cal J}_0[\phi,\wit{\phi}]$ has nontrivial
symmetry properties, in contrast to the full action
${\cal J}[\phi,\wit{\phi}]$, where these symmetries are
destroyed by the noise. We call the theory with respect to
${\cal J}_0[\phi,\wit{\phi}]$ {\em noise-free} and denote
averages taken with respect to ${\cal J}_0[\phi,\wit{\phi}]$ only
by $\langle \ldots \rangle_0$. Averages of the full
theory can then be computed by formally expanding around the
noise-free theory
\BEQ
\label{n_point3}
 \Big\langle \mathcal{O} \Big\rangle = \left\langle
 \mathcal{O} \exp(-\mathcal{J}_{\rm b}[\wit{\phi}]) \right\rangle_0
\EEQ
The noise-free theory has a Gaussian structure, if we consider the
two-component field $\Psi =\left({\phi \atop \wit{\phi}}\right)$. 
Then the factor $\exp(-{\cal J}_0[\phi,\wit{\phi}])$ can be written as 
$\exp(-\int \!\D u\D \vec{R} \D u' \D \vec{R}'\, \Psi^t \mathsf{A} \Psi)$
with an antidiagonal matrix $\mathsf{A}$.
{}From this, two important facts can be deduced:
\begin{enumerate}
   \item Wick's Theorem holds \cite{Zinn02}. That is, we can
   write the $2n$-point function as
    \BEA
     \label{wick}
     & &
     \hspace{-2.0cm}
     \Big\langle
     \phi(t_1,\vec{r}_1) \ldots
     \phi_{2n}(t_{2 n},\vec{r}_{2n})\Big\rangle_0 =       
     \sum_{\mbox{all possible  pairings} \atop \mbox{ P of {$1,2,\ldots,2n$}}}
     \\ & & \hspace{-1.0cm} \Big\langle \phi_{P_{1}}
     (t_{P_{1}},\vec{r}_{P_{1}})
     \phi_{P_{2}}(t_{P_{2}},\vec{r}_{P_{2}}) \Big\rangle_0
     \ldots \Big\langle
     \phi_{P_{2n-1}}(t_{P_{2n-1}},\vec{r}_{P_{2n-1}})
     \phi_{P_{2n}}(t_{P_{2n}},\vec{r}_{P_{2n}}) \Big\rangle_0
     \nonumber
    \EEA
   \item We have the statement
     \begin{equation}
         \label{selection_rule2}
	 \left\langle \, \underbrace{\phi \ldots \phi}_n \: 
	 \underbrace{\wit{\phi} \ldots \wit{\phi}}_m \, 
	 \right\rangle_0 = 0
     \end{equation}
     unless $n = m$. This is due to the antidiagonal
     structure of $\mathsf{A}$ and can be seen by performing the Gaussian
     integral (see for instance
     \cite{Taeuber}, chapter 4) explicitly and taking care
     of the fact, that the inverse matrix $\mathsf{A}^{-1}$
     is antidiagonal again. Formally, eq.~(\ref{selection_rule2}) is 
     analogous to the Bargman superselection rule which is used in 
     Schr\"odinger-invariant theories with $z=2$ \cite{Picone04}. 
\end{enumerate}
With these tools at hand we cam demonstrate, quite analogous to the
non-conserved case \cite{Picone04} an exact reduction of any average to an
average computed only with the noiseless theory. For example, for the two-time response function, we have
\BEQ
R(t,s) = \left\langle \phi(t) \wit{\phi}(s)\right\rangle =
\left\langle \phi(t) \wit{\phi}(s) e^{-{\cal J}_b[\wit{\phi}]} \right\rangle_0 
= \left\langle \phi(t) \wit{\phi}(s) \right\rangle_0
\EEQ
In going to the last line, we have expanded the exponential and use that because of Wick's theorem any $2n$-point function can rewritten as a sum over a 
product of $n$ two-time functions which in turn are determined by the Bargman
rules. From the special structure of $J_b[\wit{\phi}]$ it follows that only
a single term of the entire series remains. As a consequence, the two-time
response function does not depend explicitly on the noise. 
A similar result holds for the 
correlation function, see section~5. We shall now first consider the noise-free 
theory and find the two-point function from the dynamical symmetry. 
Afterwards, we shall show that the two-point functions of the full noisy 
theory can be reconstructed from this case.

\subsection{The response function of the noise-free theory}

First, we consider the linear response function of the order-parameter with
respect to an external magnetic field $h$
\BEQ
\label{def_r}
R(t,s;\vec{x}-\vec{y}) := \left.\frac{\delta \langle
\phi(t,\vec{x}) \rangle}{\delta h(s,\vec{y})}\right|_{h=0}
\EEQ
As already mentioned in the previous section,
we choose here, and in contrast to \cite{Roethlein06}, 
a perturbation respecting the
conservation law, which means that we have to add the term
$\nabla_{\vec{R}}^2 h$ to the Langevin equation
(\ref{start_equation}) or the term $\int \!\D\vec{R} \D u
\,h \nabla_{\vec{R}}^2 \wit{\phi} $ to the action
(\ref{action}). Then it is easy to see that the response
function (\ref{def_r}) is given by
\begin{equation}
\label{def_r2}
R(t,s;\vec{x}-\vec{y}) =  \Big\langle \phi(t,\vec{x})
\nabla_{\vec{y}}^2 \wit{\phi}(s,\vec{y}) \Big\rangle 
\end{equation}
We make the important assumption that the field 
$\phi(t,\vec{r})$, characterised by the parameters 
$\lambda$ and $\gamma$ and the scaling dimension $x$, is quasiprimary. 
As suggested in \cite{Henkel02,Picone04} we consider also the response field
$\wit{\phi}(t,\vec{r})$ to be quasiprimary, with
parameters $\wit{\lambda}=-\lambda$ and $\wit{\gamma}=-\gamma$ but the same 
scaling dimension $\wit{x}=x$.

We can concentrate on the model with $v=0$ for the following
reason: Suppose $\phi(\vec{r},t)$ is a solution of equation 
(\ref{start_equation}) with $\eta = 0$. Then we define
\begin{equation}
  \label{gauge_transform}
  \Psi(t,\vec{r}) := \exp \left(\frac{1}{16 \lambda} \int_0^t \!\D\tau v(\tau)
  \nabla^2_{\vec{r}}\right) \phi(t,\vec{r}).
\end{equation}
$\Psi(t,\vec{r})$ fulfils the following dynamical
equation 
\begin{equation}
\partial_t \Psi(t,\vec{r}) = \frac{1}{16 \lambda} \nabla^4_{\vec{r}}
\Psi(t,\vec{r}),
\end{equation}
which is the same equation with $v = 0$. If suffices thus to
consider the problem with $v = 0$ and then to apply the
inverse of the gauge transformation (\ref{gauge_transform}) for
treating the case $v \neq 0$. In this way, the breaking of time-translation
invariance is implemented. 

\subsection{The case $v = 0$}

This case is relevant for $d>4$ in the spherical model and
for any $d$ in the Mullins-Herring model. 
We compute (\ref{def_r2}) for the noise-free theory,
taking translation-invariance into account. Then the response function 
is given by
\BEQ
\label{nabla_r}
R_0(t,s;\vec{r}) = \nabla^2_{\vec{r}}
F^{(2)}(t,s,\vec{r})
\EEQ
where the two-point function $F^{(2)}(t,s,\vec{r})$ has been
computed in the last section. We obtain, with the scaling variable
$u=|\vec{r}|(t-s)^{-1/4}$  
\BEA
R_0(t,s;\vec{r}) &=& (t-s)^{-{(x+1)}/{2} }
\left( u^{1-d} \partial_u (u^{d-1} \partial_u)
\right) \sum_{s \in \mathcal{E}'} {c}_s {\phi}^{(s)}(u)
\nonumber \\
\label{lsiresult_response}
&=& (t-s)^{-{(x+1)}/{2}}
\sum_{s \in \mathcal{E}''} \wht{c}_s \wht{\phi}^{(s)}(u)
\EEA
where $\wht{c}_s$ are constants, $\mathcal{E}''$ the set of
admissible values for $s$ given below, 
and the solutions $\wht{\phi}^{(s)}$ are
given by
\BEQ
\wht{\phi}(u)^{(s)} = u^{s-6} \sum_{\ell=0}^\infty
\wht{b}_{\ell}^{(s)}u^{4 \ell}
\EEQ
with the coefficients
\BEQ
\wht{b}_{\ell}^{(s)} = \left( -\lambda \right)^{\ell}
\frac{\Gamma(\frac{s}{2}+1) \Gamma(\frac{s}{2}+\frac{d}{2})
}{\Gamma(\frac{s+d}{4} - \frac{1}{2} -
\frac{\gamma}{\lambda})} 
\cdot
\frac{\Gamma(\ell + \frac{s+d}{4} - \frac{1}{2} - \frac{\gamma}{
\lambda})}{\Gamma(2 \ell + \frac{s}{2} -2) \Gamma(2 \ell +
\frac{s}{2} + \frac{d}{2}-3)}
\EEQ
A priori, $s$ could take the values $s \in \{2,4,2-d,4-d
\} \setminus \{-2,-4,\ldots \}$ as derived in appendix~B.
However, certain values of $s$ have to be
excluded, since the solution has to be regular for $u
\rightarrow 0$ and has to vanish for $u \rightarrow \infty$.
By inspection of the coefficients $\wht{b}_{\ell}^{(s)}$ one
finds, in a similar way as done at the end of appendix B:
\BEQ
\mathcal{E}'' := \left\{ \begin{array}{lcl} \{2,4\} &
                                \mbox{if} & d \geq 4 \\
                            \{2,4,2-d\} & \mbox{if} & 2 <
			    d < 4\\
                            \{2,4,2-d,4-d\}& \mbox{if} & d
			    \leq 2 
                           \end{array} \right.
\EEQ
In the specific example of the spherical model in turns out that 
$\frac{\gamma}{\lambda} = \frac{1}{2}$. 
Then the solution with $s = 4-d \in \mathcal{E}''$ disappears
and is no longer admissible for $ d \leq 2$.

For completeness, we rewrite the solutions $\phi^{(s)}(u)$ as
hypergeometric functions for the admissible $s$, suppressing some
constant prefactors which can be absorbed into the constants
$\wht{c}_s$.
\BEA
\wht{\phi}^{(2)}(u) &=& {_1F}_3
\left(1+\frac{d}{4}-\frac{\gamma}{\lambda};
\frac{1}{2}, \frac{d}{4},\frac{1}{2}+\frac{d}{4}; -
\frac{\lambda u^4}{16}\right) \\
\wht{\phi}^{(4)}(u) &=& \left( -\frac{\lambda u^4}{16}
\right)^{{1}/{2}} {_1F}_3 \left(\frac{3}{2}+\frac{d}{4}-
\frac{\gamma}{\lambda};
\frac{3}{2},\frac{1}{2}+ \frac{d}{4},1 +\frac{d}{4}; -
\frac{\lambda u^4}{16}\right) \\
\wht{\phi}^{(2-d)}(u) &=& \left( -\frac{\lambda u^4}{16}
\right)^{1-{d}/{4}} {_1F}_3 \left(2 -
\frac{\gamma}{\lambda};
\frac{3}{2}-\frac{d}{4},2-\frac{d}{4},\frac{3}{2}; -
\frac{\lambda u^4}{16}\right) \\
\wht{\phi}^{(4-d)}(u) &=& \left(-\frac{\lambda u^4}{16}
\right)^{{(2-d)}/{4}} {_1F}_3 \left(\frac{3}{2}-
\frac{\gamma}{\lambda};
1-\frac{d}{4},\frac{3}{2}- \frac{d}{4},\frac{1}{2}; -
\frac{\lambda u^4}{16}\right) 
\EEA
The constants $\wht{c}_s$ are not completely arbitrary for the
case $\lambda < 0$, but have
to be arranged so that $\wht{\phi}(u) \rightarrow 0$ for $u
\rightarrow \infty$. From \cite{Wright1,Wright2}, one knows
the asymptotic behaviour of the hypergeometric functions. In general,
there is an infinite series of terms growing exponentially with $u$, together
with terms falling off algebraically.  The leading term of the exponentially
growing series can be eliminated by imposing the following 
condition on the coefficients
\BEA
\label{condition_coef}
& &
c_2 \frac{\Gamma(\frac{1}{2}) \Gamma(\frac{d}{4})
\Gamma(\frac{1}{2} + \frac{d}{4})}{\Gamma(1+\frac{d}{4}-
\frac{\gamma}{\lambda})} + c_4
\frac{\Gamma(\frac{3}{2}) \Gamma(\frac{1}{2}+\frac{d}{4})
\Gamma(1 + \frac{d}{4})}{\Gamma(\frac{3}{2}+\frac{d}{4}
-\frac{\gamma}{\lambda})}\\
&+&  c_{2-d}
\frac{\Gamma(\frac{3}{2}-\frac{d}{4}) 
\Gamma(2-\frac{d}{4}) \Gamma(\frac{3}{2})}{\Gamma(2-\frac{\gamma}{\lambda})} +
c_{4-d} \frac{\Gamma(1-\frac{d}{4}) \Gamma(\frac{3}{2}-\frac{d}{4})
\Gamma(\frac{1}{2})}{\Gamma(\frac{3}{2}-\frac{\gamma}{\lambda})} 
= 0 \nonumber
\EEA
Here, some of the constants $\wht{c}_s$ might have to be set to
zero, if the corresponding value of $s$ is not admissible.
Indeed, the condition (\ref{condition_coef}) is sufficient to cancel
the {\em entire} exponentially growing series and the remaining
part decreases algebraically, but we shall not prove this here.
The most important case for us is $c_{2-d} = c_{4-d} = 0$.
In this case (\ref{condition_coef}) implies the relation
$c_2 = -\frac{\Gamma\left(\frac{d}{4}
+ \frac{1}{2}\right)}{\Gamma\left(\frac{d}{4}+
\frac{3}{2}-\frac{\gamma}{\lambda }\right)} \frac{c_4}{2} $ and it
is easy to show that (see appendix B for the details)
\BEA
\label{integral_rep}
R_0(t,s;\vec{r}) &=& \nabla_{\vec{r}}^2 F^{(2)}(t,s,\vec{r})
\nonumber \\
& &\hspace{-2.5cm} = r_0 (t-s)^{-{(x+1)}/{2}} \int
\frac{\D\vec{k}}{(2\pi)^d} \left( k^2 \right)^{2-\frac{2
\gamma}{\lambda}} \exp \left(-\II
\frac{\vec{k}\cdot\vec{r}}{(t-s)^{{1}/{4}}}\right) \exp\left(-k^4\right)
\EEA
This prediction of LSI with $z = 4$ is perfectly consistent
with the exact results (\ref{response_result}) and
(\ref{result_response}) of the conserved spherical model for the case $d
> 4$ if we set $\frac{\gamma}{\lambda} = \frac{1}{2}$ and $x
= \frac{d}{2}$. The case $d < 4$ in the spherical model 
(where we have $v \neq 0$)
will be treated in the next subsection.

\subsection{The case $v \neq 0$}

In this case, the response function is given by
\BEQ
\label{response_vn0}
R(t,s;\vec{r}) = \exp( \xi_{t,s}
\nabla_{\vec{r}}^2)\nabla_{\vec{r}}^2  F^{(2)}(t,s,\vec{r})
\EEQ
where we have defined $\xi_{t,s} :=
-\frac{1}{16 \lambda} \int_s^t \!\D \tau\,
v(\tau)$ using the fact that $\wit{\phi}$ is
characterised by the parameters $-\lambda$ and $-\gamma$. Therefore,
remembering $u=|\vec{r}|(t-s)^{-1/4}$ 
\BEQ
\label{result_rv2}
\hspace{-1.0cm}
R_0(t,s,\vec{r}) = r_0 (t-s)^{-{(x+1)}/{2}}
\sum_{n=0}^\infty
\frac{(\xi_{t,s}(t-s)^{-\frac{1}{2}})^n}{n!}
(\nabla_{u}^2)^{n} \sum_{s \in \mathcal{E}''}\wht{c}_s
\wht{\phi}^{(s)}(u)
\EEQ
where $\nabla_{u}^2 = u^{1-d} \partial_u
(u^{d-1} \partial_u)$ and the $\wht{\phi}^{(s)}(u)$ have been 
given in the preceding subsection.
We now make the following assumption,
following the idea suggested in \cite{Picone04}: If we want
to have scaling behaviour, we need that
\BEQ
\int_0^t \!\D \tau\, v(\tau) \stackrel{t \rightarrow
\infty}{\sim} \kappa_0 t^{\digamma}
\EEQ
at least for long times, which implies  $\xi_{t,s}
\sim -\frac{1}{16 \lambda } \kappa_0 (t^{\digamma} -s^{\digamma})$. 
Evaluating the
expression (\ref{result_rv2}), we find
\BEA
\label{full_r}
& &R_0(t,s;\vec{r}) = r_0 (t-s)^{-{(x+1)}/{2}}
\sum_{s \in \mathcal{E}''}\wht{c}_s \sum_{\ell,n=0}^\infty
\frac{(\xi_{t,s}(t-s)^{-\frac{1}{2}})^n}{n!} 2^{2 n}
\wht{b}_{\ell}^{(s)} \nonumber \\ & &\times 
\frac{\Gamma(2\ell + \frac{s}{2} -2) \Gamma(2 \ell
+ \frac{s}{2}-3 + \frac{d}{2})}{\Gamma(2 \ell + \frac{s}{2} - 2
-n) \Gamma(2 \ell + \frac{s}{2} - n - 3 + \frac{d}{2})} 
u^{4 \ell + s - 2 n - 6} 
\EEA
Now we introduce $u = |\vec{r}|(t-s)^{-1/4}$ and
$\xi_{t,s} \sim -\frac{1}{16 \lambda} \kappa_0 (t^\digamma -
s^\digamma)$ into this expression. By changing the summation
variables from $n$ and $\ell$ to $k := 4 \ell - 2n +s -6$ and $\ell$ one
can then read off the dynamical exponent, which is given by
\BEQ
z = \frac{2}{\digamma}.
\EEQ
The critical exponent $z$ is therefore determined by the
behaviour of the potential $v(\tau)$. However, as we shall
see later when treating the correlation function, only a
value of $\digamma = \frac{1}{2}$ leads to scaling
behaviour of the correlation function. Nevertheless we  keep
$\digamma$ arbitrary for now and  proceed with the autoresponse
function $R(t,s)=R(t,s;\vec{0})$, which can be obtained from (\ref{full_r}) by
setting $u=0$. All nonvanishing terms have to satisfy
the condition $4 \ell + s - 2 n -6 = 0$, which excludes in
particular odd values of $s$. Therefore only the
contributions from the values $s = 2$ and $s = 4$ remain
and the result for $R(t,s)$ is
\BEQ
\label{final_result_r}
R_0(t,s) = (t-s)^{-{(x+1)}/{2}}\left[\tilde{c}_2 g^{(2)}(t,s)+ \tilde{c}_4
g^{(4)}(t,s\right].
\EEQ
$\tilde{c}_2$ and $\tilde{c}_4$ are parameters and the
expressions $g^{(2)}(t,s)$ and $g^{(4)}(t,s)$ are given by
\BEA
g^{(2)}(t,s) &=&
{_1F}_1\left(1+\frac{d}{4}-\frac{\gamma}{\lambda};
\frac{1}{2}; -4 \lambda \frac{\xi_{t,s}^2}{t-s}\right)\nonumber \\
g^{(4)}(t,s) &=& (-4 \lambda \xi_{t,s}^2
(t-s)^{-1})^{\frac{1}{2}}{_1F}_1 \left( \frac{3}{2} +
\frac{d}{4} - \frac{\gamma}{\lambda}; \frac{3}{2}; -4
\lambda \frac{\xi_{t,s}^2}{t-s}\right) 
\EEA
Note that we can write 
\BEQ
\label{expression}
\frac{\xi_{t,s}^2}{t-s} = s^{2 \digamma-1}
\left( \frac{(y^\digamma-1)^2}{y-1}\right)
\EEQ
with $y = t/s$. For future extension it is instructive to consider 
the different
asymptotic behaviour implied by $\digamma$, which we carry out in appendix C.
Here we check only that our theory is in line with the
exact results of the conserved spherical model as derived in section~2. 
If we choose $c_2 = -\frac{\Gamma\left(\frac{d}{4}
+ \frac{1}{2}\right)}{\Gamma\left(\frac{d}{4}+
\frac{3}{2}-\frac{\gamma}{\lambda }\right)} \frac{c_4}{2} $ 
and $c_{2-d} = c_{4-d} = 0$, we have seen in the last subsection 
that the expression
${\nabla_{\vec{r}}}^2 F^{(2)}(t,s,\vec{r})$ can be written
as an integral (see (\ref{integral_rep})).  
On this integral representation, we apply formula (\ref{response_vn0}). We
also recall the fact that $ \exp(\xi_{t,s} \nabla_{\vec{r}}^2)
e^{\II k r} = e^{\II k r - \xi_{t,s} k^2}$. It then 
follows from a straightforward computation 
that the result (\ref{response_result}) is reproduced
correctly for $\kappa_0 = -\mathfrak{g}_d$. 

\section{Response and correlation function in the noisy theory}

\subsection{The response function}

We use the decomposition (\ref{splitup})
and expand around the noise-free theory. Because of
(\ref{selection_rule2}), we find that the response
function is equal to the noise-free response function, derived in
the last section
\BEQ
R(t,s;\vec{r}) = R_0(t,s;\vec{r})
\EEQ
Therefore, we can take over the results already
discussed in the last section. In particular, we can conclude that
the form of the two-time response function in the critical spherical model 
with conserved order-parameter agrees with the prediction of local 
scale-invariance. 

\subsection{The correlation function}

For the correlation function, the following terms remain
(as usual $\vec{r} = \vec{x}-\vec{y}$)
\BEA
\label{correlation1}
C(t,s;\vec{r}) &=& -\frac{T}{16 \lambda} \int \!\D u
\D\vec{R}\, \Big\langle
\phi(t,\vec{x}) \phi(s,\vec{y}) \wit{\phi}(u,\vec{R})
\nabla_{\vec{R}}^2 \wit{\phi}(u,\vec{R}) \Big\rangle_0 \nonumber \\
& &+ \frac{a_0}{2} \int \!\D\vec{R}\, \Big\langle \phi(t,\vec{x})
\phi(s,\vec{y}) \wit{\phi}^2(0,\vec{R}) \Big\rangle_0
\EEA
where we have assumed uncorrelated initial conditions, that
is
\BEQ
\langle \phi(\vec{R},0) \phi(\vec{R}',0) \rangle = a_0
\delta(\vec{R}-\vec{R}')
\EEQ
We denote the first term on the right-hand side of (\ref{correlation1})
by $C_1(t,s;\vec{r})$ and the second term by
$C_2(t,s;\vec{r})$.
Unfortunately, we do not have an expression for the
three- and four-point functions. However, we can use Wick's
theorem and the Bargman superselection rule, which leads to
\BEA
\label{c1}
& & \hspace{-1.5cm}C_1(t,s;\vec{r}) 
= -\frac{T}{16 \lambda} \int_0^s \!\D u\int \!\D\vec{R} \,
\Big\langle \phi(t,\vec{x}) \wit{\phi}(u,\vec{R}) \Big\rangle_0
\nabla_{\vec{R}}^2 \Big\langle \phi(s,\vec{y})
\wit{\phi}(u,\vec{R}) \Big\rangle_0 \nonumber \\
&&  \hspace{0.5cm}-  \frac{T}{16 \lambda} \int_0^s \!\D u\int \!\D \vec{R} \,
\nabla_{\vec{R}}^2 \Big\langle \phi(t,\vec{x})
\wit{\phi}(u,\vec{R}) \Big\rangle_0
\Big\langle \phi(s,\vec{y}) \wit{\phi}(u,\vec{R}) \Big\rangle_0 
\EEA
and
\BEQ
\label{c2}
C_2(t,s;\vec{r}) = a_0 \int \!\D\vec{R}\, \Big\langle
\phi(t,\vec{x})
\wit{\phi}(0,\vec{R}) \Big\rangle_0
\Big\langle \phi(s,\vec{y}) \wit{\phi}(0,\vec{R}) \Big\rangle_0
\EEQ
Under the integrals, we always find two sorts of factors:
The two-point function $F^{(2)}(t,s;\vec{r}) = \langle
\phi(t,\vec{x}) \wit{\phi}(s,\vec{y}) \rangle$ was computed in
appendix~A and reads ($ \vec{r} = \vec{x}-\vec{y}$ and $u =
|\vec{r}| (t-s)^{-1/4}$)
\BEQ
\label{twopoint_correlation}
F^{(2)}(t,s;\vec{r}) = (t-s)^{-{x}/{2}} \sum_{s
\in \mathcal{E}'} c_s \phi^{(s)}(u)
\EEQ
and the response function $R_0(t,s,\vec{r}) = \langle
\phi(t,\vec{x})(\nabla_{\vec{y}})^2 \wit{\phi}(s,\vec{y}) \rangle$ as
computed in the previous section
\BEQ
\label{response_correlation}
R_0(t,s;\vec{r}) = (t-s)^{-{(x+1)}/{2}} \sum_{s
\in \mathcal{E}''} \wht{c}_s \wht{\phi}^{(s)}(u)
\EEQ
When we introduce the expressions (\ref{twopoint_correlation}) and
(\ref{response_correlation}) with the appropriate arguments
into (\ref{c1}) and (\ref{c2}) we get the most general form
of the correlation function fixed by LSI with $z = 4$. \\ 

We now show, that this result is compatible with the
exactly known result for $C(t,s;\vec{r})$ of the conserved spherical model
as derived in section~2. From the response function, we have already seen that
we need to make the choice $c_{2-d} = c_{4-d} = 0$ and $c_2 = -\frac{\Gamma\left(\frac{d}{4}
+ \frac{1}{2}\right)}{\Gamma\left(\frac{d}{4}+
\frac{3}{2}-\frac{\gamma}{\lambda }\right)} \frac{c_4}{2} $. 
Then we have the integral representation (\ref{integral_rep}) for
$R(t,s;\vec{r})$ to which one can still apply the gauge
transform (\ref{gauge_transform}). 
For $F^{(2)}(t,s,\vec{r})$ one can find in
the same way an integral representation, so that we have the
following two expressions:
\BEA
\label{integral_rep2}
F^{(2)}(t,s;\vec{r}) &=& (t-s)^{-{x}/{2}} \int
\frac{\D\vec{k}}{(2\pi)^d} \left( k^2 \right)^{1-\frac{2
\gamma}{\lambda}} \exp \left(-
\frac{\II \vec{k}\cdot\vec{r}}{(t-s)^{{1}/{4}}}\right) \nonumber \\
& & \times  \exp\left(-k^4+ \frac{k^2}{16 \lambda} \kappa_0
\left( \frac{t^\digamma - s^\digamma}{(t-s)^{{1}/{2}}}
\right)\right) \\
R_0(t,s;\vec{r}) &=& (t-s)^{-{(x+1)}/{2}} \int
\frac{\D\vec{k}}{(2\pi)^d} \left( k^2 \right)^{2-\frac{2
\gamma}{\lambda}} \exp \left(- 
\frac{\II \vec{k}\cdot\vec{r}}{(t-s)^{{1}/{4}}}\right)
\nonumber \\
& & \times  \exp\left(-k^4+ \frac{k^2}{16 \lambda} \kappa_0
\left( \frac{t^\digamma - s^\digamma}{(t-s)^{{1}/{2}}} \right)\right) 
\EEA
Using these representations, we obtain
\BEA
& & \hspace{-1.8cm} C_1(t,s;\vec{r}) = -\frac{T}{8
\gamma} s^{\frac{d}{4}-\frac{2 \gamma}{\lambda}-x +
\frac{3}{2}} \int_0^1 \!\D \theta
(y-\theta)^{\frac{d}{4}+\frac{1}{2}-\frac{\gamma}{\lambda}-\frac{x}{2}}
(1-\theta)^{\frac{d}{4}+\frac{1}{2}-\frac{\gamma}{\lambda}-\frac{x}{2}} \\
& & \times (y+1-2 \theta)^{-\frac{d}{4}-2+\frac{2
\gamma}{\lambda}}
\int \!\frac{\D\vec{k}}{(2 \pi)^d}\: 
\exp \Big(- \frac{\II \vec{k}\cdot\vec{r}}
{(t+s-2 s \, \theta)^{\frac{1}{4}}}\Big) (k^2)^{3-\frac{4
\gamma}{\lambda}} \nonumber
 \\
& & \times \exp\left(-{k}^4 + \frac{{k}^2}{16 \lambda}\kappa_0
 \frac{t^\digamma + s^\digamma-2(s \, \theta)^\digamma}{(t+s-2
 s \, \theta)^{{1}/{2}}}\right) \nonumber \\
& &\hspace{-1.8cm} C_2(t,s;\vec{r}) = a_0
(t+s)^{-\frac{d}{4}-1+ \frac{2\gamma}{\lambda}} \left(t
 \,s\right)^{\frac{d}{4}+\frac{1}{2}-\frac{\gamma}{\lambda}-\frac{x}{2}} 
 \int \!\frac{\D\vec{k}}{(2 \pi)^d}\: \exp \Big(-
\frac{\II \vec{k} \cdot \vec{r}}
{(t+s)^{{1}/{4}}}\Big)({k}^2)^{2-\frac{4\gamma}{\lambda}} 
\nonumber \\
 & & \times \exp\left(-{k}^4 + \frac{{k}^2}{16 \lambda}\kappa_0
 \frac{t^\digamma + s^\digamma}{(t+s)^{{1}/{2}}}\right) 
\EEA
where we have used the fact that the integration over
$\vec{R}$ gives a delta function. In order to compare
with the exact results eqs.~(\ref{a1},\ref{a2}), we see directly that we
have the equalities $A_1(t,s;\vec{r}) = C_2(t,s;\vec{r})$ and
$A_2(t,s;\vec{r}) = C_1(t,s;\vec{r})$ 
for $x = \frac{d}{2}$, $\lambda = -\frac{1}{16}$ and
$\frac{\gamma}{\lambda} = \frac{1}{2}$ and $\digamma =\frac{1}{2}$.
Our symmetry-based approach has thus reproduced all the
results from section~2, up to an identification of parameters.

\section{Conclusion}

Current studies on non-equilibrium systems often concentrate on the scaling
aspects, since then universality may be used to justify the analysis of rather
artificial-looking models in order to obtain physical insight. In this work, 
we have been exploring the idea that dynamical scaling might be generalisable
to a larger algebraic structure of local scale-transformations. This kind
of prediction is most conveniently tested through the form of the two-time
response functions and, since the form of the autoresponse function contains
the dynamic exponent $z$ only in combination with other exponents, the scaling
form of the space-time response function will provide non-trivial information
about how to construct local scale-transformation beyond the case $z=2$ which
by now is rather well understood. Since values of $z$ quite distinct from two
can be obtained for a conserved order-parameter, this motivated our choice to
study the kinetics of such systems. The spherical model, quenched to $T=T_c$
from a fully disordered initial state and the
Mullins-Herring model with conserved noise are useful first
tests, since their
scaling behaviour is non-trivial, yet the models do remain analytically
treatable. 

By looking for generalised local scale-transformation, it might appear at first
sight that the non-invariance of a stochastic Langevin equation under such
transformations would make such a programme futile. However, generalizing a
technique from Schr\"odinger-invariance (which applies for example to
phase-ordering kinetics with a non-conserved order-parameter \cite{Picone04}),
we have shown that also in the conserved case one can decompose the Langevin
equation into a deterministic part which does admit larger symmetries and a
noise part which breaks them. This decomposition is such that all interesting
averages can be exactly reduced to averages calculable from the deterministic
part of the theory. We have shown that this decomposition can be carried out
in the critical spherical model and did confirm 
that the predictions obtained for the
two-time response and correlation functions are fully compatible with the
exactly known results in the spherical model. Together with the Mullins-Herring
growth model, these are the first analytically solved examples which confirm
LSI for a dynamical exponent $z=4$.\footnote{Another example
of a system with a dynamical exponent far from $2$ which
appears to be compatible with LSI is the bond-diluted 2D
Ising-model quenched to $T < T_c$ \cite{Henkel06b}.} In table 1 we
collect the values of the exponents of the conserved
spherical model and the conserved and the non-conserved 
Mullins-Herring models. For
comparison we also list some of the values for the exponents
of the bond-diluted  2D Ising-model, for which it was shown
that the autoresponse function is in agreement with LSI \cite{Henkel06b}. 

That confirmation was possible because the deterministic part of the Langevin
equation is still a linear equation. Numerical simulations in models where
this is no longer so will inform us to what extent LSI with $z\ne 2$ can be 
confirmed in a more general context. At the same time, it will be
necessary to derive a generalisation of the Bargman superselection rules in
order to have a model-independent justification of the decomposition of the
Langevin equation. Work along these lines is in progress. 

\begin{table}[t]
\[
\hspace{-2.0cm}
\begin{array}{||c||c|c|c|c|c||} \hline \hline 
\mbox{model} & a & b & \lambda_R & \lambda_C & \mbox{Reference}  \\
\hline \hline
\mbox{cons. spherical} & d/4-1/2 & d/4 -1/2 
& d+2 & d+2 & \cite{Calabrese05} ~\&~ \mbox{this work}
\\ \hline
\mbox{cons. Mullins-Herring} & d/4-1/2 & d/4 -1/2 
& d+2 & d+2 & \cite{Calabrese05} ~\&~ \mbox{this work}
\\ \hline
\mbox{non-cons. Mullins-Herring} & d/4-1 & d/4-1 & d & d & \cite{Roethlein06}
\\ \hline
\mbox{diluted Ising}& 0.24(2) &  & 1.32(4) & 1.280(20) & \cite{Henkel06b} \\ 
\hline \hline
\end{array}
\]
\caption{\label{table1} Exponents $a,b,\lambda_R,\lambda_C$
of the conserved spherical model and the Mullins-Herring
model at criticality and the bond-diluted 2D Ising model
below the critical temperature. The latter model 
is defined by the hamiltonian $\mathcal{H} = -\sum_{(i,j)}
J_{ij} \sigma_i \sigma_j$ with $\sigma_i = \pm 1$ and 
random variables $J_{ij}$, which are equally distributed over
the intervall $[1-\epsilon/2,1+\epsilon/2]$ with $ 0 \leq
\epsilon \leq 2$. For the given values of the exponents of
this model the
temperature has the value $T = 1.0$ and 
the parameter $\epsilon$ has the value $\epsilon = 2.0$, see 
\cite{Henkel06b} for details. For all models, the dynamical
exponent is $z = 4$.
}
\end{table}

\appsection{A}{The two-point function}

In this appendix, we compute the two-point function. We do
the computation for $d=2$ for convenience, but the case for
arbitrary dimension will be obvious. We write
\BEQ
F^{(2)}(t_1,t_2, r_1^x,r_1^y,r_2^x,r_2^y) = \Big\langle
\phi_1(t_1,r_1^x,r_1^y) \phi_2(t_2, r_2^x,r_2^y)
\Big\rangle
\EEQ
where $r_i^x$ denotes the $x$-component of the coordinate
vector $\vec{r}_i$ and $r_i^y$ the y-component.
Invariance under $X_{-1},Y^{(x)}_{-\frac{1}{4}}$ 
and $Y^{(y)}_{-\frac{1}{4}}$ gives us
\BEQ
F^{(2)} = \tilde{F}(t,r^{x},r^{y}),
\EEQ
where we have defined:
\BEQ
r^{x} := r_1^{x} - r_2^{x}, \;\;\; r^{y} := r_1^{y}
- r_2^{y}, \;\;\; t := t_1 -t_2 
\EEQ
Next we take the rotation generator
\BEQ
R^{(x,y)} = r_x \partial_{r_y} - r_y \partial_{r_x}
\EEQ
and find
\BEQ
(r^{x} \partial_{r^y} - r^y \partial_{r^x})
\tilde{F}(t,r^x,r^y) = 0
\EEQ
{}From this we can conclude that
\BEQ
\tilde{F} = G(t,r) \qquad \mbox{with} \;\;\; r =
\sqrt{r_x^2 + r_y^2} = |\vec{r}|
\EEQ
Now we use invariance under $X_0$. It requires
\BEQ
\left(t_1 \partial_{t_1} + t_2 \partial_{t_2} + \frac{1}{4}
\Big(r_1^x \partial_{r_1^x} + r_1^y \partial_{r_1^y} + r_2^x
\partial_{r_2^x} + r_2^y \partial_{r_2^y}\Big) + {x}\right) F = 0
\EEQ
with ${x} = (x_1+x_2)/{4}$. This can be transformed into
\BEQ
\label{condition_x0}
\left(t \partial_t + \frac{1}{4} r \partial_r +
x\right) G = 0
\EEQ
Here we have used the facts
\BEA
\partial_{r_1^x} = \frac{r_x}{\sqrt{r_x^2 + r_y^ 2}}
\partial_r &,\quad&
\partial_{r_1^y} = \frac{r_y}{\sqrt{r_x^2 + r_y^ 2}}
\partial_r \nonumber \\
\partial_{r_2^x} = -\frac{r_x}{\sqrt{r_x^2 + r_y^ 2}}
\partial_r &,\quad&
\partial_{r_2^y} = -\frac{r_x}{\sqrt{r_x^2 + r_y^ 2}}
\partial_r 
\EEA
Before we proceed, we also note the fact that, when applied
to $G(t,r)$, we have the identity
\BEQ
\label{nablasquare}
\partial^2_{r_1^x} + \partial^2_{r_1^y} = \partial^2_{r_2^x}
+ \partial^2_{r_2^y} = \partial^2_r + \frac{1}{r} \partial_r
= \frac{1}{r} \partial_r (r \partial_r)
\EEQ
This is just the operator $\nabla^2_{\vec{r}}$ in polar coordinates
without angular part. For arbitrary dimension $d$ we must
use
\BEQ
\partial^2_{r^{x_1}} + \ldots +\partial^2_{r^{x_d}} =
\frac{1}{r^{d-1}}\cdot \partial_r (r^{d-1} \partial_r) 
\EEQ
instead of (\ref{nablasquare}).
Turning to the operators $Y_{\frac{3}{4}}^{(x_i)}$, we
require 
\BEQ
\label{selection_rule}
\lambda = \lambda_1 = -\lambda_2,
\qquad \gamma = \gamma_1 = - \gamma_2.
\EEQ
Then invariance under $Y_{\frac{3}{4}}^{(x_i)}$ eventually
leads to
\BEQ
\label{condition_y22}
\hspace{-2.5cm}
\left(t \partial_{r^{(x_i)}} + 4 \lambda 
r^{(x_i)} \left(\frac{1}{r^{d-1}}
\partial_{r} (r^{d-1} \partial_r)\right)^{-1} - 16 \gamma \partial_{r^{(x_i)}} 
\left(\frac{1}{r^{d-1}} \partial_{r} 
(r^{d-1} \partial_r)\right)^{-2} \right) G = 0
\EEQ
or, after multiplication with $\frac{r}{r^{(x_i)}}$
\BEA
\label{condition_y2}
\hspace{-2.0cm}
\left(t \partial_r + 4 \lambda r
\left(\frac{1}{r^{d-1}}
\partial_r (r^{d-1} \partial_r)\right)^{-1}
 -16 \gamma \partial_r \left(\frac{1}{r^{d-1}}
\partial_r (r^{d-1} \partial_r)\right)^{-2} \right) G =
0
\EEA
independently of the direction.
Lastly, invariance under $X_1$ yields after a
straightforward but slightly lengthy calculation
\BEA
\label{condition_x1}
& &\hspace{-2.0truecm}
\left( t^2 \partial_t + \frac{t}{2} r \partial_r + 2
t x_1 + 2
t_2 X_0 + \sum_{i=1}^d \frac{r^{(x_i)}}{2}
Y_{\frac{3}{4}}^{(x_i)} +
\lambda r^2 \left(\frac{1}{r^{d-1}} \partial_r (r^{d-1}
\partial_r)\right) \right. \nonumber \\ 
&-& \left. 4 \gamma r \partial_r
\left(\frac{1}{r^{d-1}} \partial_r (r^{d-1}
\partial_r)\right)^{-2}\right)G = 0
\EEA
Equations (\ref{condition_x0}),(\ref{condition_y2}) and
(\ref{condition_x1}) have to be solved. We multiply
(\ref{condition_x0}) by $-t$ and (\ref{condition_y2}) by $-
{r}/{4} $ and add them to (\ref{condition_x1}).
The resulting equation is satisfied provided that the
condition
\BEQ
x_1 = x_2
\EEQ
for the scaling dimensions holds. On the other hand
(\ref{condition_x0}) is solved by
\BEQ
G(t,r) = \delta_{x_1,x_2}
t^{-(x_1+x_2)/4} \phi(u), \qquad \mbox{with} \qquad u = r
t^{-{1}/{4}}.
\EEQ
{}From (\ref{condition_y2}), the scaling function $\phi(u)$ 
satisfies the following equation
\BEQ
\hspace{-2.0cm}
 \left( \partial_u + 4 \lambda u \left(
 \frac{1}{u^{d-1}} \partial_u \left( u^{d-1} \partial_u
 \right)\right)^{-1} - 16 \gamma
 \partial_u \left( \frac{1}{u^{d-1}} \partial_u \left(
 u^{d-1} \partial_u \right)\right)^{-2} \right)
 \phi(u) = 0
\EEQ
This we rewrite using the shorthands $\wht{a} := 4
\lambda $ and $\wht{b} := - 16 \gamma$
\BEA
\label{scaling_function}
\hspace{-1.5cm}
\left(\partial_u \left( \frac{1}{u^{d-1}} \partial_u \left(
u^{d-1} \partial_u \right) \right)^2 + \wht{a} u \left(
\frac{1}{u^{d-1}} \partial_u \left( u^{d-1} \partial_u
\right) \right) + \wht{b} \partial_u\right) \wit{\phi}(u) & = &
0
\EEA
with 
\BEQ
\label{sc1:sc2}
\wit{\phi}(u) := \left( \frac{1}{u^{d-1}} \partial_u
\left(u^{d-1} \partial_u \right) \right)^{-2} \phi(u)
\EEQ
We note the following property
\BEQ
\left(\frac{1}{u^{d-1}} \partial_u \left( u^{d-1} \partial_u
\right) \right) u^{\alpha} = \alpha (\alpha+d-2)
u^{\alpha-2}
\EEQ
 We make a similar ansatz as for the case $d=1$
\cite{Roethlein06}
\BEQ
\wit{\phi}(u) = \sum_{n=0}^\infty c_n u^{n+s} \qquad c_0
\neq 0
\EEQ
Introducing this into (\ref{scaling_function}), we obtain on
the one hand, because of $c_0 \neq 0$
\BEQ
s \in \{0,2,4,2-d,4-d\} =: \mathcal{E}.
\EEQ
On the other hand, we get a recursion relation for $c_n$
\BEQ
\label{recursion}
c_n = -\frac{\wht{a} (n+s+d-6) +
\wht{b}}{(n+s)(n+s+d-2)(n+s-2)(n+s+d-4)} c_{n-4}
\EEQ
Third, we get an additional assumption on $s$, namely
\BEQ
\label{add_cond}
s \neq -2,-4,\ldots, \quad \mbox{and} \quad s \neq -d, -d-2,\ldots
\EEQ
If we did not have this, we would encounter an equation of
the form $c_n \cdot 0 + c_{n-4} \xi = 0$ with $\xi \neq 0$,
which would imply $c_{n-4} = 0$. However, this would be in
contradiction to $c_0 \neq 0$.
The recursion (\ref{recursion}) can be solved by standard methods (compare
for instance \cite{Henkel02}). One can show that for $s \in
\mathcal{E}$, one may set $c_1 = c_2 = c_3 = 0$
\footnote{This can in fact be deduced, except for some
special cases in low dimensions. The latter are however covered
by the different values of $s$} 
and solve the recursion starting from $c_0$. 
We also set  $\tilde{b}_{\ell} := c_{4\ell}$ and obtain as
the solution of the recursion:
\BEQ
\tilde{b}_{\ell} = \left( -\frac{\wht{a}}{4} \right)^{\ell}
\frac{\Gamma(\frac{s}{2}+1) \Gamma(\frac{s}{2}+\frac{d}{2})
}{\Gamma(\frac{s+d}{4} - \frac{1}{2} +
\frac{\wht{b}}{4 \wht{a}})}
\cdot
\frac{\Gamma(\ell + \frac{s+d}{4} - \frac{1}{2} + \frac{\wht{b}}{4
\wht{a}})}{\Gamma(2 \ell + \frac{s}{2}+1) \Gamma(2 \ell + \frac{s}{2} +
\frac{d}{2})}.
\EEQ
{}From this we can deduce the form of the scaling function
$\phi(u)$ by equation (\ref{sc1:sc2}).
The final result for the scaling function is therefore
\BEQ
\phi(u) = \sum_{s \in \mathcal{E}} c_s \phi^{(s)}(u)
\EEQ
where $c_s$ are free parameters and $\phi^{(s)}(u)$ are
given by (substituting the original expressions for $\wht{a}$
and $\wht{b}$).
\BEQ
 \phi^{(s)}(u) = \sum_{\ell = 0}^\infty
 b_{\ell}^{(s)} u^{4 \ell + s-4}
\EEQ
with
\BEQ
\label{app_coefficients}
\hspace{-1.0cm}
b_{\ell}^{(s)} =  2^4 \left(-\lambda \right)^\ell
\frac{\Gamma(\frac{s}{2}+1)
\Gamma(\frac{s}{2}+\frac{d}{2})}
{\Gamma(\frac{s+d}{4} - \frac{1}{2} -
\frac{\gamma}{\lambda})}
\cdot \frac{
\Gamma(\ell + \frac{s+d}{4} - \frac{1}{2} - \frac{\gamma}{\lambda})}
{\Gamma(2 \ell + \frac{s}{2}-1)
\Gamma(2 \ell + \frac{s}{2} + \frac{d}{2}-2)
}
\EEQ
This kind of function has been considered for instance in
\cite{Wright1,Wright2} and can in fact be rewritten as
generalised hypergeometric functions with the result, up to normalisation
\BEA
\phi^{(2)}(u) &\sim& \left( -\frac{\lambda u^4}{16} \right)^{\frac{1}{2}}
{_1F}_3\left(\frac{d}{4}+1-\frac{\gamma}{\lambda};
\frac{3}{2},\frac{1}{2}+\frac{d}{4},\frac{d}{4}+1;-\frac{\lambda
u^4}{16} \right) \nonumber \\
\phi^{(4)}(u) &\sim&
{_1F}_3\left(\frac{d}{4}+\frac{1}{2}-\frac{\gamma}{\lambda};
\frac{1}{2},\frac{d}{4},\frac{d}{4}+\frac{1}{2};-\frac{\lambda
u^4}{16} \right)  \\
\phi^{(2-d)}(u) &\sim& 
\left( -\frac{\lambda u^4}{16}
\right)^{\frac{1}{2}-\frac{d}{4}}
{_1F}_3\left(1-\frac{\gamma}{\lambda};
1-\frac{d}{4},\frac{3}{2}-\frac{d}{4},\frac{1}{2};-\frac{\lambda
u^4}{16} \right) \nonumber \\
\phi^{(4-d)}(u) &\sim& 
\left( -\frac{\lambda u^4}{16} \right)^{1-\frac{d}{4}}
{_1F}_3\left(\frac{d}{4}+\frac{3}{2}-\frac{\gamma}{\lambda};
\frac{3}{2}-\frac{d}{4},2-\frac{d}{4},\frac{3}{2};-\frac{\lambda
u^4}{16} \right) \nonumber
\EEA
Here the $s$-dependent prefactors have been dropped, as they
can be absorbed into the parameters $c_s$. Note however that
for certain values of $s$ and $\gamma/\lambda$ these factors
might be zero so that the corresponding solution vanishes,
see below for the free-field case.
The coefficients as given in (\ref{app_coefficients}) will be used in the main
text to determine the response function. Let us collect
however some facts about the solutions $\phi^{(s)}(u)$, as
this might be relevant for models with non-conserving
perturbations.
\begin{itemize}
   \item To be physically acceptable, the solutions have to
   be regular for $u \rightarrow 0$ and vanishing for $u
   \rightarrow \infty$. The solutions 
   belonging to $s \in \{0,2,4\}$ are indeed
   regular for $u \rightarrow 0$. However, the solution
   for $s \in \{4-d,2-d\}$  are not regular for $d > 4$ and
   have to be dropped for this case (for even $d$ they where
   anyway excluded for $d > 4$). For the same reason, the
   solution for $s = 2-d$ has to be abandoned for $d > 2$.
   \\
   
   Furthermore, we have $\tilde{b}_{\ell}^{(4)} =
   \tilde{b}^{(0)}_{\ell+1}$ and $\tilde{b}_0^{(0)} = 0$, so that,
   up to a constant, the solution $\phi^{(4)}(u)$ equals
   $\phi^{(0)}(u)$. Therefore, we drop the solution belonging
   to $s = 0$. The general solution is then
   \BEQ
   \phi(u) = \sum_{s \in \mathcal{E}'} c_s \phi^{(s)}(u)
   \EEQ
   with 
   \BEQ
   \mathcal{E}' := \left\{ \begin{array}{lcl} \{2,4\} &
                                \mbox{if} & d > 4 \\
                            \{2,4,4-d\} & \mbox{if} & 2 <
			    d \leq 4\\
                            \{2,4,2-d,4-d\}& \mbox{if} & d
			    \leq 2 
                           \end{array} \right.
   \EEQ
   
   \item For free-field theories, there is
   $\frac{\gamma}{\lambda} = \frac{1}{2}$. This makes the solution
   for $s = 4-d$ vanish (see (\ref{app_coefficients})).
   \item The scaling function must vanish for $u \rightarrow
   \infty$. Hence, not all of the
   coefficients $c_s$ need to be independent, but they have to be
   arranged in such a way as to achieve this asymptotic behaviour. 
\end{itemize}

\appsection{B}{Calculation of an integral}
\label{app_integral}

We outline briefly how to obtain the integral
\BEQ
\label{integral}
I_{\vec{r}}^{(\alpha,\beta)} := \int_{\mathbb{R}^d} \!\D \vec{k}\; 
(\vec{k}^2)^{1+\beta} e^{\II \vec{k}\cdot \vec{r}-\alpha (\vec{k}^2)^2} 
\EEQ
We choose spherical coordinates, that is: 
\BEA
k_1 & = & k \cos(\theta_1) \nonumber \\
k_2 & = & k \sin(\theta_1) \cos(\theta_2) \nonumber \\
\vdots & & \qquad  \vdots \\
k_{d-1} & = & k \sin(\theta_1) \sin(\theta_2)\cdot \ldots
\cdot \sin(\theta_{d-2}) \cos(\theta_{d-1}) \nonumber \\
k_{d} & = & k \sin(\theta_1) \sin(\theta_2)\cdot \ldots
\cdot \sin(\theta_{d-2}) \sin(\theta_{d-1}) \nonumber
\EEA
where $k = |\vec{k}|$ and $0 \leq \theta_1 <
\pi,\ldots,0\leq \theta_{d-2}< \pi,0 \leq \theta_{d-1} < 2 \pi$.
Furthermore we rotate the system, so that we have
$\vec{k}\cdot \vec{r} = k r \cos(\theta_1)$ with $r = |\vec{r}|$.
The Jacobian is given by
\BEQ
\frac{\partial(k_1,\ldots,k_n) }{
\partial(k,\theta_1,\ldots,\theta_{d-1}) } = k^{d-1} \sin^{d-2}(\theta_1)
\sin^{d-3}(\theta_2)\cdot \ldots \cdot \sin(\theta_{d-2})
\EEQ
If we change to the new coordinates in (\ref{integral}), the
angles $\theta_2,\ldots,\theta_{d-1}$ can be integrated out,
yielding a prefactor $\mathcal{A}'$, independent of $\vec{r},\alpha$ and
$\beta$, which we determine later.
Therefore
\BEQ
I_{\vec{r}}^{(\alpha,\beta)} 
= \mathcal{A}' \int_0^\infty \!\D k\, k^{d+1+2\beta} e^{-\alpha k^4}
\int_0^\pi \!\D \theta_1\: e^{\II k r \cos(\theta_1)}
\sin^{d-2}(\theta_1)
\EEQ
The inner integral gives $\Gamma(\frac{d}{2}-\frac{1}{2})
\Gamma(\frac{1}{2}) (\II k r/2)^{1-\frac{d}{2}}
I_{\frac{d}{2}-1}(\II k r) $, where $I_\nu(z)$ is a modified Bessel
function \cite{Grad}, for which we use the series expansion
$I_\nu(z) = \sum_{n=0}^\infty ({n!
\Gamma(\nu + n + 1)})^{-1} (z/2)^{\nu + 2 n}$. Then we perform the 
remaining integral over $k$. This leaves us with the expression
\BEQ
\label{integral_sol}
I_{\vec{r}}^{(\alpha,\beta)} = \mathcal{A}\,
\alpha^{-\frac{d}{4}-\frac{1}{2}-\frac{\beta}{2}} 
\sum_{n=0}^\infty \frac{\left(-{r^2}/{(4 \sqrt{\alpha})}\right)^n
}{n! \Gamma(\frac{d}{2}+n)} \,
\Gamma \left(\frac{d}{4}+\frac{1}{2}+\frac{\beta}{2} +
\frac{n}{2}\right).
\EEQ
with a new factor $\mathcal{A}$ which is also independent of
$\vec{r},\alpha$ and $\beta$.
This factor can now easily be determined by
computing $I_{\vec{0}}^{(\alpha,-1)}$ directly and then
comparing with (\ref{integral_sol}). This yields
$\mathcal{A} = {\pi^{d/2}}/{2}$.
Splitting the sum into even and odd terms and using for the
Gamma functions the fact that $\Gamma(2 z) =
\pi^{-\frac{1}{2}} 2^{2 z -1} \Gamma(z)
\Gamma(z+\frac{1}{2}) $, we can rewrite this
result in terms of hypergeometric functions and obtain
\BEA
& & \hspace{-1.5cm}
I_{\vec{r}}^{(\alpha,\beta)} 
= \int_{\mathbb{R}^d} 
\!\D \vec{k}\; (\vec{k}^2)^{1+\beta}e^{\II \vec{k}{r} -\alpha 
(\vec{k}^2)^2}  = \mathcal{G} \times \left[
{_1F}_3 \left(\frac{d}{4}+\frac{1}{2} +
\frac{\beta}{2}; \frac{1}{2}, \frac{d}{4},
\frac{d}{4}+\frac{1}{2}; \frac{r^4}{256 \,\alpha}\right)
\nonumber \right. \\ && \hspace{-1.5cm} \left.
-\frac{8 \, \Gamma\left(\frac{d}{4}+1+
\frac{\beta}{2}\right)}{d \, \Gamma \left(
\frac{d}{4}+\frac{1}{2}+\frac{\beta}{2}\right)} 
\left(\frac{r^2}{16 \sqrt{\alpha}} \right)
{_1F}_3\left( \frac{d}{4}+1+\frac{\beta}{2}; \frac{3}{2},
\frac{d}{4}+\frac{1}{2}, \frac{d}{4} + 1; \frac{r^4}{256 \,\alpha
}\right)\right] \nonumber 
\EEA
with the expression
\BEQ
\mathcal{G} = 
\left(\frac{\pi}{2}\right)^{\frac{d}{2}} \sqrt{\pi}
\alpha^{-\frac{d}{4}-\frac{1}{2}-\frac{\beta}{2}}
\frac{\Gamma(\frac{d}{4}+\frac{1}{2}+\frac{\beta}{2})}
{\Gamma(\frac{d}{4}+\frac{1}{2})
\Gamma(\frac{d}{4})}
\EEQ

\appsection{C}{The response function in the scaling regime}

We reconsider equation (\ref{final_result_r}) and consider the
different cases implied by $\digamma$.
\begin{itemize}
   \item \underline{$\digamma = \frac{1}{2}
   \Leftrightarrow z = 4$ }. This case in fact
   corresponds to the conserved spherical model.  
   In this case both $g^{(2)}(t,s)$ and $g^{(4)}(t,s)$
   depend only on the ratio $t/s$ and the argument
   (\ref{expression}) of the hypergeometric functions tends
   to one for $y \rightarrow \infty$. Therefore, in the
   scaling regime we obtain:
   \BEQ
   \label{exp_f12}
   a = \frac{x}{2}-\frac{1}{2}, \qquad \mbox{and} \quad
   \lambda_R = 2 x + 2
   \EEQ
   \item \underline{$\digamma < \frac{1}{2}
   \Leftrightarrow z > 4$}: In this case, the
   arguments of the hypergeometric functions tend to zero in
   the scaling limit and we obtain
   \BEQ
   R_0(t,s) = s^{-\frac{x}{2}-\frac{1}{2}}
   (y-1)^{-\frac{x}{2}-\frac{1}{2}} \left(\tilde{c}_2 +
   \tilde{c}_4 \sqrt{-\frac{4 \lambda \kappa_0}{4}} s^{\digamma-\frac{1}{2}}
   \frac{y^\digamma-1}{\sqrt{y-1}} \right)
   \EEQ
   if $\tilde{c}_2 \neq 0$, this leads to
   \BEQ
   R_0(t,s) = \tilde{c}_2 s^{-\frac{x}{2}-\frac{1}{2}}
   (y-1)^{-\frac{x}{2}-\frac{1}{2}}
   \EEQ
   and the same values (\ref{exp_f12})for $a$ and
   $\lambda_R$ as before. If $
   \tilde{c}_2 = 0$, we have
   \BEQ
   R_0(t,s) = \tilde{c}_4 s^{-\frac{x}{2}+\digamma-1}
   (y-1)^{-\frac{x}{2}-1}(y^\digamma -1)
   \EEQ
   which means for the exponents:
   \BEQ
   a = \frac{x}{2}-\digamma, \qquad \frac{\lambda_R}{z} =
   \frac{x}{2} + 1 - \digamma
   \EEQ
   \item \underline{$\digamma >
   \frac{1}{2}\Leftrightarrow z < 4$:} 
   If $\lambda < 0$, the
   coefficients $\tilde{c}_2$ and $\tilde{c}_4$
   have to be arranged in a way, that the
   exponentially growing parts of the hypergeometric
   functions cancel. More precisely, the condition 
   \BEQ
   \tilde{c}_4
   \frac{\Gamma(\frac{3}{2})}
   {\Gamma(\frac{3}{2}+\frac{d}{4}-\frac{\gamma}{\lambda})}
   + \tilde{c}_2
   \frac{\Gamma(\frac{1}{2})}{\Gamma(1+\frac{d}{4} -
   \frac{\gamma}{\lambda})} = 0
   \EEQ
   has to be satisfied. In this case we find 
   \BEQ
   \label{auto1}
   R(t,s) = \wht{r}_0 s^{-\frac{x}{2}-\frac{1}{2}-(2 \digamma
   -1)(1+\frac{d}{4}-\frac{\gamma}{\lambda})} 
   \left( \frac{(y^\digamma -1)^2}{(y-1)}\right)
   ^{-(1+\frac{d}{4}-\frac{\gamma}{\lambda})}
   \EEQ
   where $\wht{r}_0$ is a normalisation constant. This leads
   to the following critical exponents:
   \BEA
   \label{auto1_exp}
   a &=& \frac{x}{z} - \frac{1}{2} + (2 \digamma
   -1)\left(1+\frac{d}{4}-\frac{\gamma}{\lambda}\right) \\
   \lambda_R/z &=& \frac{x}{2} + \frac{1}{2} + (2 \digamma
   -1)\left(1+\frac{d}{4}-\frac{\gamma}{\lambda}\right)
   \EEA
   If $\lambda > 0$ there is no condition
   on $\tilde{c}_2$ and $\tilde{c}_4$, as the exponential
   contribution decreases rapidly. In this case expressions
   (\ref{auto1}) and (\ref{auto1_exp}) remain valid (only
   the constant $\wht{r}_0$ may change) provided
   $\tilde{c}_2 \Gamma(\frac{1}{2}) \neq \tilde{c}_4
   \Gamma(\frac{3}{2})$. If the latter condition is not
   fulfilled, the result is
   \BEQ
   R(t,s) = \wht{r}_0 s^{-\frac{x}{2}-\frac{1}{2}-(2 \digamma
   -1)(\frac{d}{4}-\frac{\gamma}{\lambda})} 
  \left( \frac{(y^\digamma -1)}{(y-1)} \right)
  ^{-2(\frac{d}{4}-\frac{\gamma}{\lambda})}
   \EEQ
   and
   \BEA
   a &=& \frac{ x}{z} - \frac{1}{2} + (2 \digamma
   -1)\left(\frac{d}{4}-\frac{\gamma}{\lambda}\right) \\
   \lambda_R/z &=& \frac{x}{2} + \frac{1}{2} + (2 \digamma
   -1)\left(\frac{d}{4}-\frac{\gamma}{\lambda}\right)
   \EEA
\end{itemize}
In all cases we find the relation
\BEQ
\lambda_R = z (a + 1)
\EEQ

\appsection{D}{Conserved random walk}

Conceivably the most simple ageing system is the well-known Brownian
particle, described by a Langevin equation only containing the noise and 
the external perturbing field \cite{Cugl94}. Here we extend
their idea to a conserved order-parameter $\phi$ and consider the following 
model, described by the Langevin equation
\BEQ
\partial_t \phi(t,\vec{x}) = - \nabla_{\vec{x}}^2 h(t,\vec{x})
+ \eta(t,\vec{x})
\EEQ
where the $\eta$ is the same kind of noise (\ref{noise_corr}) as in the main
text and $h$ is an external field. 
The solution of the model is promptly found and we obtain, 
where as usual $\vec{r} = \vec{x}-\vec{y}$
\BEA
R(t,s;\vec{r}) &=& - (\nabla_{\vec{r}}^2 \delta(\vec{r})) \Theta(t-s) \\
C(t,s;\vec{r}) &=& C(0,0;\vec{r}) -2 T (\nabla_{\vec{r}}^2
\delta(\vec{r})) \mbox{min}(t,s)
\EEA
where the initial term $C(0,0;\vec{r})$ can be dropped in
the scaling limit. The limit fluctuation-dissipation ratio is $X_{\infty}=1/2$. As in the 
non-conserved case \cite{Cugl94}, an equilibrium state is never reached. 

The difference to the non-conserved case is that the space-dependence is
described by a second derivative of a Delta-function instead of a simple
delta-function. In order to better understand the meaning of this result, 
we return to a lattice discretization of the model. 
Then $\delta(\vec{r})$ goes over to a Kronecker delta 
$\delta_{\vec{r},\vec{0}}$ and the second derivative becomes a discrete
difference. We illustrate the result in one space dimension 
\BEA
R(t,s;{r}) &=& \Theta(t-s)\left( 2 \delta_{{r},0} -
\delta_{{r}+1,0} - \delta_{{r}-1,0} \right) \\
C(t,s;{r}) &=& 2 T \mbox{min}(t,s)\left( 2 \delta_{{r},0} -
\delta_{{r}+1,0} - \delta_{{r}-1,0} \right)
\EEA
Then, for $t>s$, $R(t,s;0)=2$ is a local maximum, while $R(t,s;\pm 1)=-1$
would correspond to local minima. This is qualitatively similar to the form
of $R(t,s;\vec{r})$ seen in figure~\ref{figure1} for the conserved spherical 
model.  

\noindent 
{\bf Acknowledgements:}
We thank A. J. Bray, C. Godr{\`e}che and S. Majumdar for discussions and the 
Isaac Newton Institute for hospitality, where part of this work 
was done. We acknowledge the support by the Deutsche Forschungsgemeinschaft
through grant no. PL 323/2.
This work was supported by the franco-german binational
programme PROCOPE.


\end{document}